\documentclass[aps,twocolumn,prd,nofootinbib,showpacs]{revtex4}
\usepackage[pdftex,bookmarks]{hyperref}

\usepackage{amsfonts}
\usepackage{amsmath,amssymb}
\usepackage[T1]{fontenc}
\usepackage{mathrsfs}
\usepackage{epsfig,epstopdf}

\usepackage[absolute]{textpos}

\usepackage{graphics,wrapfig,times}
\usepackage{graphicx}

\TPGrid[40mm,40mm]{15}{25}  

 \makeatletter
 \newcommand{\pback}[1]{{
   \let\@rrow=\leftarrowfill
   \mathchoice{\AIN@stemPullBack{#1}{\@rrow}}{\AIN@stemPullBack{#1}{\@rrow}}
     {\AIN@indxPullBack{#1}{\@rrow}}{\AIN@indxPullBack{#1}{\@rrow}}}
   \vphantom{#1}}

 \newcommand{\AIN@stemPullBack}[2]{
   \vtop{\mathsurround=0pt
   \ialign{##\crcr$\textstyle{#1}\strut$\crcr
     \noalign{\kern-0.4ex\nointerlineskip}{\tiny#2}\crcr}}}

 \newcommand{\AIN@indxPullBack}[2]{
   \vtop{\mathsurround=0pt
   \ialign{##\crcr\hfil$\scriptstyle{#1}$\hfil\crcr
     \noalign{\kern+0.4ex\nointerlineskip}{\tiny#2}\crcr}}}

\def\bar{\overline}
\def\be{\begin{equation}}
\def\ee{\end{equation}}
\def\bea{\begin{eqnarray}}
\def\eea{\end{eqnarray}}
\def\ba{\begin{array}}
\def\ea{\end{array}}

\def\={\hateq}
\def\puto#1{\rlap{\raise.5ex\hbox{\char'27}}{#1}}

\newcommand{\nn}{\nonumber}
\newcommand{\half}{\frac{1}{2}}
\def\eth{\text{\dh}}
\def\thorn{\text{\th}}
\def\a{\alpha}
\def\b{\beta}
\def\c{\gamma}
\def\d{\delta}

\def\IH{\triangle}

\def\heq{\hat{=}}

 \def\tt2{{\tilde{2}}}

\def\.{\cdot}
\def\D{{\cal D}}

\def\L{{\cal L}}

\def\M{{\mathscr M}}
\def\kl{\kappa_{(\ell)}}

\def\Re{{\rm Re}}
\def\Im{{\rm Im}}
\def\l{\ell}

\def\be{\begin{equation}}
\def\ee{\end{equation}}
\def\bea{\begin{eqnarray}}
\def\eea{\end{eqnarray}}
\def\ba{\begin{array}}
\def\ea{\end{array}}

\def\R{{\mbox{\rm$\mbox{I}\!\mbox{R}$}}}

\def\up{\stackrel}

\newcommand{\eqhat}{\mathrel{\widehat\mathalpha{=}}}
\def\={\eqhat}

\begin{document}

\title{Gravitational radiations of generic isolated horizons and non-rotating dynamical horizons from asymptotic expansions}

\author{Yu-Huei Wu}
\email{yhwu@phy.ncu.edu.tw, yuhueiwu@hotmail.com}
\affiliation{1. Center for Mathematics and Theoretical Physics, National Central University. \\
2. Department of Physics, National Central University,\\
     Chungli, 320, Taiwan.}

\author{Chih-Hung Wang}
\email{chwang@phy.ncu.edu.tw}
\affiliation{Department of Physics, National Central University,\\
     Chungli, 320, Taiwan.}

\begin{abstract}

Instead of using a three dimensional analysis on quasi-local
horizons, we adopt a four dimensional asymptotic expansion
analysis to study the next order contributions from the
nonlinearity of general relativity. From the similarity between null
infinity and horizons, the proper reference
frames are chosen  from the compatible constant spinors for an observer to measure the energy-momentum and flux
near quasi-local horizons. In particular, we focus on the
similarity of Bondi-Sachs gravitational radiation for the
quasi-local horizons and compare our results to Ashtekar-Kirshnan flux formular. The quasi-local energy momentum and flux of
generic isolated horizons and non-rotating dynamical horizons are
discussed in this paper.

\end{abstract}
\date{\today}
\pacs{95.85.Sz,04.70.-s,11.30.-j}
\maketitle

\section{Introduction}


The boundary of a black hole is defined as  \textit{a region of no
escape of null rays}, which leads to an event horizon definition. However, the definition of the event horizon cannot give a realistic description of how a black hole grows since it is too
global. The event horizon can be located only after an observer knows the
global structure of space-time. Hence, the generalization of the event horizon to a \textit{quasi-local} definition of horizons provides \textit{an observer a possibility to detect horizons}.
Ashtekar \textit{et al} \cite{Ashtekar99b} proposed a notion of isolated horizons, which is quasi-locally defined, to describe equilibrium states of black holes. Unlike Killing horizons or stationary event horizons, they do not require any space-time Killing fields or exclude radiations outside horizons. If
gravitational collapse occurs, the final stage of black
holes is isolated and in the equilibrium state therefore will not
radiate any more. However, there should still have gravitational or
matter field radiations outside black holes. The definitions of generic isolated horizons yield a theoretical framework to describe these situations.
%
%

In order to clarify the idea of 'isolation', Ashtekar \textit{et al} \cite{Ashtekar99b} started from the most general definition of isolated
horizons called \textit{non-expanding horizons} (NEHs). It requires the intrinsic (degenerate) metric $q_{ab}$ on horizons to be time independent, i.e., $\L_\l q_{ab} \heq 0$, where $\hat{=}$ represents equal on horizon. Hence, $\l$, which is a null normal to horizon, can be considered as a Killing vector field of intrinsic horizon geometry. If one
further requires the extrinsic curvature (the rotation 1-form) to
be time independent, it then leads to the definition of \textit{weakly
isolated horizons} (WIHs).
In WIHs, the black hole zeroth and first laws holds. By using a freedom of rescaling $\l$, it has been verified that there exists a specific $\l$ which can reduce NEHs to WIHs.  The most restricted generic isolated horizons are called \textit{isolated horizons} (IHs), which require the full derivative operator $\D$ induced by space-time connection $\nabla$ to be time independent.

It is expected that black holes are rarely in equilibrium in
Nature. By using generic isolated horizons as a basis, the ideas
can be generalized to \textit{dynamical horizons} (DHs) definition by considering a
space-like hypersurface rather than a null hypersurface in the
NEH definition. The horizon geometry of
dynamical horizon is time dependent and it allows a quantitative
relation between the growth of the horizon area and the flux of
energy and angular momentum across it \cite{Ashtekar02}. Ref \cite{Ashtekar02} adopted a 2+1 decomposition on a three dimensional
space-like surface and the Cauchy data on the DHs must
satisfy the scalar and vector constraints. Moreover,
the relations between changes of the horizon
area and energy fluxes cross the DHs were obtained and these fluxes has also been proved to be positive. However, this kind of approach
does not tell us what is the gravitational free data near horizon
when considering the full four dimensional space-time.

In contrast to Ashtekar \textit{et al}'s \cite{Ashtekar99b, Ashtekar02} three dimensional analysis, our works use 4-dimensional asymptotic expansions to study the neighborhoods of generic IHs and DHs. Since asymptotic expansion has been used to study gravitational radiations near the null infinity \cite{NU}, it offers a useful scheme to analyze gravitational radiations approaching another boundary of space-time, horizons. We first set up a null frame with certain gauge choices
near quasi-local horizons and then expand Newman-Penrose (NP) coefficients, Weyl, and Ricci curvature with respect to radius. Their fall-off can be determined from NP equations, Bianchi equations, and exact solutions, e.g., the Kerr solution (see Appendix \ref{Kerr}) and the Vaidya solution (see Appendix \ref{qlVaidya}).    This approach allows one to see the next order contributions from
the nonlinearity of the full theory for the quasi-local horizons.
The asymptotic expansions for the null infinity and the horizon are quite different geometrically. As we
approach to null infinity, we consider an asymptotically flat
space-time, however, the approach near horizons is \textit{not
necessarily asymptotically flat}. For the null infinity, we take
the incoming null vector $n^a$ as a generator of null infinity to
generate different cuts with respect to different times, say $u$, and the
outgoing null vector $\l^a$
is parameterized by using affine parameter $r$. Moreover, $\l_a$ is chosen as the gradient of the surface of a constant time $u$. For generic IHs, we take the
outgoing null vector $\l^a$ as the generator of horizons to
generate different cross sections with respect to different times, say $v$. The ingoing null vector $n^a$ is chosen as
a tangent vector of ingoing null rays and also being parameterized by
affine parameter $r$. The $n_a$ is the gradient of the surface of a constant time $v$.


The main purpose of this paper is to calculate
the amount of mass-energy fluxes crossing or near generic IHs and non-rotating DHs.
Although
there is no well-defined gravitational energy density in general relativity (GR), it does have well-defined mass or energy associated to a closed two-surface, i.e.,
quasi-locally. Without using Hamiltonian or 2+1 decomposition on these quasi-local horizons, we can use a quasi-local energy formula based on
spinor fields to define the mass of a black hole. How can one measure the quasi-local energy-momentum when she or he is near black holes? Our strategy is to use the Nester-Witten 2-form together with similar concept of Bramson's frame alignment \cite{Bramson75a}. The Nester-Witten 2-form is commonly used to prove the positivity of energy and can be defined quasi-locally. It has been proved to be positive either for the stationary space-time or the space-time with radiation, i.e, the ADM mass or Bondi mass. Therefore we use it for a near horizon region which can allow radiation. However, what is \textit{a good reference frame} for a strong gravitating field such as a black hole? To tackle this problem, we adopt the notion of frame alignment to find asymptotically constant spinors, which was done by Bramson for null infinity, and apply this framework on horizons. After obtaining the compatible constant spinors, we use them to define the quasi-local energy-momentum and flux near horizons. Since we know the changes of the constant spinors with respect to time, it is possible for us to see how does the energy flux cross horizons.


The plan of this paper is as follows. Section \ref{IH-DH}
gives a review of generic IHs and DHs. Their definitions and properties are discussed in terms of the NP coefficients.
In Section \ref{Asym-exp}, we present a detail calculation of asymptotic expansions for IHs and non-rotating DHs. The asymptotic expansion provides a useful scheme to analyze the space-time geometry not only on quasi-lcoal horizons but also their neighborhoods. In Section \ref{conspinor},  we apply the Bramson's frame alignment to IHs and non-rotating DHs and then find the compatible constant spinors. Thus, the good reference frame can be constructed from these constant spinors. On each cross section of quasi-local horizons, we find that constant spinors will satisfy the \textit{Dogan-Mason's holomorphic condition}.
In Section \ref{energy-flux}, we obtain the quasi-local energy and flux
of the generic IHs and non-rotating DHs by using these constant spinors.
Although our approaches, which is based on gauge fixing quasi-local energy-momentum and \textit{a time related condition}, are completely different from Ashtekar \textit{et al}'s, it turns out to yield the same result as Ashtekar-Kirshnan gravitational fluxes for non-rotating DHs.

In this paper, we adopt the same notation as in the references \cite{Ashtekar99b, Ashtekar02} for describing generic IHs and DHs. However, we choose the different convention $(+---)$, which is a standard convention for the NP formalism \cite{NP} (also see \cite{Chandrasekhar}).  The Weyl tensor is completely specified by  the five complex scalars $\Psi_0,...,\Psi_4$ and the ten components of the Ricci tensor are defined in terms of the four real and three complex scalars $\Phi_{00},..., \Phi_{22}$.  Their definitions can be found in p. 43-p. 44 of  \cite{Chandrasekhar}. The necessary equations, i.e., commutation relations, NP equations and Bianchi identities, for asymptotical expansion analysis can be found in p.
45-p. 51 of \cite{Chandrasekhar}.

\section{Definitions of quasi-local horizons} \label{IH-DH}

In this section, we give a review of generic IHs and DHs proposed by Ashtekar \textit{et al} \cite{Ashtekar99b, Ashtekar02}. The definitions of the quasi-local horizons yields several conditions on NP coefficients  and these conditions are called gauge conditions throughout this paper.
\subsection{The generic isolated horizons \label{IH}}


 We start from a 4-dimensional space-time manifold $(\M, g)$ with a 3-dimensional, null
sub-manifold $(\Delta, q)$. The definition of non-expanding horizon is given as follows \cite{Ashtekar99b}:

\paragraph{Definition.} $\Delta$ is called a \textit{non-expanding
horizon} (NEH) if
 (1) $\Delta$ is diffeomorphic to the product $S \times
\R$ where $S$ is a space-like 2-sphere.
 (2) The expansion $\Theta_{(\l)}$of any null normal $\l$ to $\Delta$
vanishes, where the expansion is defined by $\Theta_{(\l)}=\half
q^{ab}\nabla_a \l_b$ with $q_{ab}$ the degenerate intrinsic metric
on $\Delta$.
 (3) Einstein field equations hold on $\Delta$ and the stress-energy tensor $T_{ab}$ is such that
$T^a_{b}\l^b$ is causal and future-directed on $\Delta$.

\smallskip

The geometry of $\Delta$ is characterized by the intrinsic metric $q$ and the induced
derivative $\D$, i.e., $\D_a =\!\pback{\nabla_a}$, on $\IH$.\footnote{
The under-arrow $\!\pback{}$ indicates the pullback of the index to $\IH$.} The
intrinsic degenerate metric $q_{ab}$ on $\IH$ has signature $(0,-,-)$. The
vectors $(\l^a, m^a, \bar m^a)$ span the tangent space to $\Delta$
with the dual co-frame given by the pull backs of $(n_a, m_a, \bar
m_a)$. The expansions of outgoing and incoming null rays are defined
by
$ \Theta_{(\l)} := \half q^{ab} \nabla_a \l_b = - \Re [\rho],$ and
$\Theta_{(n)} := \half q^{ab} \nabla_a n_b =  \Re [\mu],$
where $q^{ab}:= -m^a\bar m^b-\bar m^a m^b$ on the tangent space of
$\Delta$. The twist and shear on $\Delta$ are defined as
 $\omega^2_{twist}:= \half q^a\,_c\; q^b\,_d \; \nabla_{[a}\l_{b]} \nabla^{[c}\l^{d]}$
and
$ |\sigma_{shear}| := [\half q^a\,_c\; q^b\,_d \nabla_{(a}\l_{b)}
\nabla^{(c}\l^{d)} - \Theta_{(\l)}^2 ]^\half$, respectively.  Since $\l$ is the
null normal of the null hypersurface, it implies the twist free.
Moreover, the shear vanishes by using Raychaudhuri equation
\footnote{For the outgoing null geodesic $\l$, the
\textit{Raychaudhuri equation} yields
\bea \L_{\l} \Theta_{(\l)} = - \Theta_{(\l)}^2
-\sigma_{shear}\bar\sigma_{shear} + \omega_{twist}^2 +
{\kappa}_{(\l)} \Theta_{(\l)} -\Phi_{00}. \nn\eea} and the
dominate energy condition. Therefore, the gauge conditions on NEH
are
\bea \kappa\hat{=}0, \sigma\hat{=}0, \rho\hat{=}0.
\label{shearfreeIH}\eea
From (\ref{shearfreeIH}), there must exist \textit{a natural connection
one-form} $\omega :=\omega_a d x^a$ on $\IH$ which can be obtained
by
\be \label{omega}\D_a \l^b\hat{=}\omega_a \l^b. \ee
%
%
%
%
The \emph{surface gravity} $\kappa_{(\l)}$ is defined as
\be\label{kappa}\kappa_{(\l)}:=\omega_a \l^a 
\ee
on NEH $\Delta$ (measured by $\l$). Note that we do not have a
unique normalization for $\l$.  Under the scale transformation $
\l \mapsto f \l$, we have
$ \omega\mapsto \omega+d \ln f $ and $ \kappa_{(\l)}\mapsto f
\kappa_{(\l)}+ f \L_\l \ln f $
which leaves Eq. (\ref{omega}) and Eq. (\ref{kappa}) invariant.


From (\ref{omega}), we get
\bea     \L_\l q_{ab} \hat{=} \!\pback{\L_\l g_{ab}} \hat{=}
q_{cb} \omega_a \l^c + q_{ac}\omega_b \l^c =0\eea
for any null normal $\l$ to $\IH$.  In fact, $\l$ is an asymptotic
Killing vector field as we approach the horizon even though
\textit{the space-time metric $g_{ab}$ may not admit a Killing
vector field in the neighborhood of $\IH$.}


The energy condition in the third point of NEH definition then
further implies that $R_{ab} \l^b$ is proportional to $\l_a$
\cite{Ashtekar02}, that is
$ R_{ab} \l^a X^b\hat{=} 0, $
for any vector field $X$ tangent to $\IH$. We then have
\be  \Phi_{00} \hat{=}  \Phi_{01} \hat{=} \Phi_{10} \hat{=} 0. \ee
Since $\l$ is expansion and shear-free, it must lie along one of
the principal null directions of the Weyl tensor. From equation
(b) and (k) in p. 46 of \cite{Chandrasekhar},  we have
\be \label{weyl}  \Psi_0 \hat{=} \Psi_1 \hat{=} 0,  \ee
so $\Psi_2$ is gauge invariant, i.e., independent of the choice of
the null-tetrad vectors $(n,m,\bar{m})$, on $\IH$. For NEH, we also have
 \bea d \omega \hat{=} 2 (\Im[\Psi_2])\; ^2\epsilon \label{domega}\eea
where $^2\epsilon$ is an area 2-form. The 2-form $d \omega$
can also be written as
\bea  2\D_{[a} \omega_{b]} =  2 (\eth \pi -\bar\eth\bar\pi) m_{[a}
m_{b]}.\eea
$\Im[\Psi_2]$ plays a role of gravitational contributions to
the angular-momentum at $\IH$. Ashtekar \textit{et al} calls $\omega$ the
\textit{rotational 1-form potential} and $\Im [\Psi_2]$  the
\textit{rotational curvature scalar}. Therefore, $\pi$ vanishes
for a non-rotating $\IH$.


Using the Cartan identity $\L_v = d i_v + i_v d$ and (\ref{domega}),
the Lie derivative of $\omega$ with respect to $\l$ gives
\be\label{dkappa} \L_\ell \omega_a \hat{=} 2\Im(\Psi_2)\, \l^b\,
{}^2\!\epsilon_{ba} + \D_a (\l^b\omega_b)\ \hat{=} \D_a\kl. \ee
On NEH, the surface gravity  may not be constant. To obtain the black hole
zeroth law such that the surface gravity is constant, one may need
a further condition, i.e., $ \L_\ell \omega_a=0$,  on NEH. It
motivates the definition of \textit{weakly isolated horizon}.


\smallskip
\paragraph{Definition.} A \textit{weakly isolated horizon} (WIH) is a NEH
with an equivalence class of null normals under constant
transformation. The flow of $\l$ preserves the rotation 1-form
$\omega$
$\L_\l \omega_a\hat{=}0 $,i.e., $[\L_\l ,\D] \l \hat{=}0.$
%


From (\ref{dkappa}), the condition of WIH basically preserves the
black hole zeroth law. Because $\l$ is tangent to $\IH$, the
evolution equation is in fact a constraint. See (B21) and (B22) of
\cite{Ashtekar99b}. Therefore, given a NEH, we can select a
canonical $[\l]$ by requiring $(\IH, [\l])$ to be a WIH satisfying
\bea {\cal L}_{\l} \mu\hat{=}0\;\; \textrm{or }\;\;
\dot\mu\hat{=}0.\eea
$\IH$ generically admits a unique $[\l]$ such that the incoming
expansion is time independent. This result will establish that a
generic NEH admits a unique $[\l]$ such that $(\IH, [\l])$ is a
WIH on which the incoming expansion $\mu$ is time independent.

\smallskip
\paragraph{Definition.} A weakly isolated
horizon $(\Delta, [\l ])$ is said to be \textit{isolated horizon}
(IH) if
$
 [{\cal L}_{\l}, \D] V \hat{=} 0,
$
for all vector fields $V$ tangential to $\Delta$ and all $\l \in
[\l ]$.
\medskip

From this definition, we have  $[\L_\l, \D] \l \hat{=} 0$ and
$[\L_\l, \D] n \hat{=} 0$. The first one gives the surface gravity
to be a constant by previous argument. So $\dot \epsilon\hat{=} 0$. The
second one gives $\dot \pi \hat{=} \dot \mu \hat{=}\dot \lambda
\hat{=} 0$.

\subsection{The dynamical horizon \label{DH}}



The generic IHs are taken as the equilibrium state of the
DHs. The DHs can be foliated by
marginally trapped surface S. Therefore, the expansion of the
outgoing tetrad vanishes.

\medskip

\paragraph{Definition} A smooth, 3 dimensional, space-like sub-manifold
$H$ of space-time is said to be a \textit{dynamical horizon} (DH)
if it can be foliated by a family of closed 2-manifold such that:
(1) on each leaf, $S$, the expansion $\theta_{(\l)}$ of one null
normal $\l^a$ vanishes, (2) the expansion $\theta_{(n)}$ of the
other null normal $n^a$ is negative.

\medskip

Here, $\theta_{(\l)}:=\half\, ^{(2)}q^{ab}\nabla_a\l_b$ and $\theta_{(n)}:=\half\,^{(2)} q^{ab}\nabla_a n_b$ where $ ^{(2)} q^{ab}= -(m^a\bar m^b+m^b\bar m^a)$ is intrinsic to the cross section $S$ of $H$.  From this definition, it basically tells us a dynamical horizon is
a space-like hypersurface which is foliated by closed, marginally
trapped two surface. The requirement of the expansion of the
incoming null normal is strictly negative since we want to
study a black hole (future horizon) rather than a white hole.
Also, it implies
\bea \Re [\rho] \hat{=}0. \label{DHgc}\eea


The angular momentum associated with cross-section $S$ is
$ J_S^\psi = -\frac{1}{8\pi} \oint_S K_{ab} \psi^a R^b d S $,
where $\psi^a$ is the rotational vector field and need not be an
axial Killing field and $K_{ab}$ is the extrinsic three curvature \cite{Ashtekar02}.
$\psi^a$ can be written as linear combinations of $m^a$ and $\bar m^a$, i.e., $\psi^a = E m^a
+\bar E \bar m^a$ where $E\neq 0$.  The extrinsic three curvature $K_{ab}$ can be
decomposed into $A ^{(2)}q_{ab}  + S_{ab} + 2 W_{(a} R_{b)} + B R_a
R_b$ which has been expressed in terms of Newman-Penrose coefficients
\cite{Wu2007, WuWang-RDH}. If we choose $\l$ to be geodesic,
$n_a = \nabla_a v$ and $m, \bar m$ tangent to the two surface \cite{Wu2007, WuWang-RDH}, we
get
\bea J_S^\psi = \frac{1}{8\pi} \oint_S \Re[\pi \bar E] d S.\eea
Hence, if $\pi\hat{=}0$, it implies a non-rotating DH.

If we use the gauge conditions $\kappa\hat{=} \pi-\bar\tau \hat{=}
\pi-(\a + \bar\b)\hat{=}0$ on a rotating DH, the total flux of
Ashtekar-Krishnan \cite{Ashtekar02} \cite{WuWang-RDH} then becomes
\bea F_{\textrm{total}} = \frac{1}{4\pi} \int [ |\sigma|^2
+|\pi|^2+ \Phi_{00}] N d^3 V \label{Ash-ii} \eea
where the gravitational flux is
\bea F_{grav} = \frac{1}{4\pi}\int_{\Delta H} N (|\sigma|^2
+|\pi|^2) d^3 V. \label{ch3-grav-flux}\eea
and the matter flux of Vaidya solution is
\bea F_{\textrm{matter}} :&=& \int_H T_{ab} T^a \l^b N d^3 V
               = \frac{1}{4\pi} \int \Phi_{00} N d^3 V \eea
where we use $4 \pi T_{ab} \l^a \l^b = \Phi_{00}$.

\section{Asymptotic expansions near quasi-local horizons} \label{Asym-exp}
\subsection{Near generic isolated horizons: vacuum}

\textbf{Frame setting, gauge choice and gauge conditions}

We choose the incoming null tetrad $n_a=\nabla_a v$ to be gradient
of the null hypersurface $v=const.$ and it gives $g^{ab} v_{,a}
v_{,a} =0$. We further choose $m, \bar m$ tangent to the two
surface. These gauge choices lead to
\be\ba{lll} \nu &=& \mu-\bar\mu = \rho-\bar\rho= \c +\bar\c = \pi-\a-\bar\b=0,\\
\pi &=& \bar\tau. \label{gauge-c}\ea \ee
From the definition of NEH, the gauge conditions from eq
(\ref{shearfreeIH}) can be expanded with respect to $r'=r-r_\Delta$ as
\bea \kappa = \kappa_0 r' + O(r'^2),  \rho   =  \rho_0 r' +
O(r'^2),  \sigma = \sigma_0 r' + O(r'^2), \nn \eea
where $\rho_0: =  [\Psi^0_2- \bar\eth_0\bar\pi_0+\pi_0\bar\pi_0]$,
$\sigma_0: = [- \bar\eth_0\bar\pi_0+\pi_0\bar\pi_0]$ from
asymptotic expansions for NP equations  \textbf{(q)} and  \textbf{(p)} in p. 47  of \cite{Chandrasekhar} \footnote{From now on, we only indicate the equation and page numbers without mention the ref \cite{Chandrasekhar}.}. Further we choose $\l$ to
be flag planes parallel on NEH, therefore it leads to
\bea \epsilon-\bar\epsilon=O(r').\eea
The rest of NP coefficients
are $O(1)$.
%
%
The Weyl tensor has the fall off (refer to equation (\ref{weyl}))
\bea \Psi_0 = O(r'), \Psi_1 = O(r').\eea
 In order to preserve orthogonal relation
$ \l^a n_a=1, m^a \bar m_a =-1, \l^a m_a= n^a m_a =0,$
we can choose the tetrad as
\bea \l^a=(1, U, X^3, X^4),\;\;
          n^a=(0,-1,0,0),\;\;
          m^a=(0,0, \xi^3,\xi^4), \nn\eea
where $U, X^3, X^4$ are real and $\xi^3, \xi^4$ are complex.

We first expand NP spin coefficients, tetrad components $U, X^k,
\xi^k$ and Weyl spinors $\Psi_k$ with respect to $r'$ and
substitute them into NP equations which include the commutation relation, Ricci identities equations and the Bianchi identities equations.
By examining equation \textbf{(a)} in p. 46, it implies $U$ must be $O(r')$.
$X^{k}$ will become $O(r')$ since we can use the condition
(4.86) in \cite{Wu2007} to perform the coordinate transformation to make
$X^{0k}=0$. 

\textbf{The radial NP equations} 

 From \textbf{(n)},\textbf{(j},\textbf{(r)},\textbf{(o)},\textbf{(q)},\textbf{(p)},\textbf{(i)},\textbf{(f)} in p. 46-47 :

 \be\ba{llll}
\mu &=& \mu_0+ ({\mu_0}^2+\lambda_0\bar\lambda_0) r' + O({r'}^2) , \\
\lambda&=&  \lambda_0+ (2\mu_0\lambda_0+ 4 \lambda_0 \c_0 + \Psi^0_4 ) r' + O(r'^2 ) ,  \\
\a     &=& \a_0 +(\lambda_0(\bar\pi_0 +\b_0) + 2\c_0\a_0+\a_0\mu_0
+ \Psi^0_3
-\bar P\up{c}\nabla \c_0) r'\\
&&+ O(r'^2) , \\
\b     &=& \b_0 + (\mu_0\bar\pi_0 -\b_0 (2 \c_0 -\mu_0) +
\a_0\bar\lambda_0 -P \up{c}\nabla \c_0) r'\\
&&+ O(r'^2),
\\
\rho   &=& [\Psi^0_2- \bar\eth_0\bar\pi_0+\pi_0\bar\pi_0] r' + O(r'^2) ,\\
\sigma &=& (-P\up{c}\nabla \bar\pi_0 + 2 \b_0 \bar\pi_0)r' + O(r'^2) ,\\
\pi   &=&  \pi_0 + [ 2 (\mu_0+\c_0) \pi_0 + 2 \lambda_0  \bar\pi_0 + \Psi^0_3] r'+ O(r'^2), \\
\epsilon&=&\epsilon_0 + [ 2\a_0 \bar\pi_0 +2 \b_0\pi_0 -2 \c_0 \epsilon_0 +
 \bar\pi_0\pi_0 + \Psi^0_2 - \dot \c_0 ]   r' \\
&&+ O(r'^2) \\
\kappa &=& \kappa_0 r' + O(r'^2)
\nn\ea\ee
From (305) and (303) in p. 45:
 \be\ba{lll}
 \xi^k &=& \xi^{k0} + O(r'), \;\;
 U = 2 \epsilon_0 r' +  O(r'^2), \\
 X^k &=&  2(\pi_0\xi^{k0}+\bar\pi_0 \bar\xi^{k0})r'+ O(r'^2).
\ea\ee
From  \textbf{(e)}, \textbf{(f)},\textbf{(g)} and \textbf{(h)} in p. 46:
 \be\ba{lll}
\Psi_0 &=& \half (- \eth_0 \Psi_1^0 + 4 \bar\pi_0 \Psi_1^0 - 3 \sigma_0 \Psi_2^0) r'^2+ O(r'^3),\\
\Psi_1 &=& (- \eth_0 \Psi_2^0 + 3 \bar\pi_0 \Psi_2^0) r'+ O(r'^2),\\
\Psi_2 & =& \Psi_2^0 + (-\eth_0 \Psi_3^0 + 2 \bar\pi_0 \Psi_3^0 + 3 \mu_0 \Psi_2^0) r'+ O(r'^2),\\
\Psi_3 &=& \Psi_3^0 + (-\eth_0 \Psi_4^0 +   \bar\pi_0 \Psi_4^0 + 4
\mu_0 \Psi_3^0) r'+ O(r'^2), \nn
 \ea\ee
\textbf{The non-radial NP equations}

From \textbf{(a)},\textbf{(b)},\textbf{(c)}, \textbf{(g)},\textbf{(d)},\textbf{(e)},\textbf{(h)},\textbf{(k)},\textbf{(m)},\textbf{(l)} in p.46-47:
\be\ba{lllll}
\dot \rho_0  &=& \bar\eth_0 \kappa_0
-\bar\kappa_0\bar\pi_0 -\kappa_0\pi_0,  & \dot \sigma_0  =  \eth_0 \kappa_0,\\
%
 \dot\pi_0 &+&  \kappa_0=0,\\
 \dot\lambda_0 &=& \bar\eth_0 \pi_0-\pi_0\bar\pi_0
- 2 \lambda_0 \epsilon_0, \\
 \dot \a_0  &-&\bar P  \up{\bar c}\nabla \epsilon_0 =0, & \dot \b_0  - P  \up{c}\nabla \epsilon_0 =0,\\
 \dot \mu_0 &=& \eth_0\pi_0 +\pi_0\bar\pi_0
  -2 \mu_0 \epsilon_0 + \Psi_2^0,  \\
\eth_0 \rho_0 &-&\bar\eth_0 \sigma_0 = - \Psi_1^0, &
 \eth_0 \lambda_0 -\bar\eth_0 \mu_0 = - \Psi_3^0, \\
 \Psi_2^0 &=& \bar P  \up{\bar c}\nabla \b_0 - P
\up{c}\nabla \a_0 +   \a_0\bar\a_0  &+ \b_0\bar\b_0 - 2 \a_0\b_0,\\
  && 2 \Im \eth_0
\pi_0 = -2 \Im \Psi^0_2.
\ea\ee
From (304) , (306) in p. 45:
\be\ba{lll}
  \kappa_0 &=& -2 P \up{c}\nabla \epsilon_0, \;\; \dot P =0,\;\;
  \bar P  \up{\bar c}\nabla \ln P = \b_0 -\bar
 \a_0.
 \ea\ee
From \textbf{(a)},\textbf{(b)},\textbf{(c)}, \textbf{(d)} in p. 49:
\be\ba{lll}
\dot \Psi_1^0
 &=&  2\epsilon_0 \Psi^0_1- 3 \kappa_0 \Psi_2^0, \;\;
 \dot \Psi_2^0
 = -2\kappa_0 \Psi^0_3,\\
 \dot \Psi_3^0
 &-& \bar\eth_0 \Psi_2^0 = 3 \pi_0 \Psi_2^0 -2 \epsilon_0
 \Psi_3^0 -\kappa_0 \Psi^0_4, \\
 \dot \Psi_4^0
 &-& \bar\eth_0 \Psi_3^0 = -3 \lambda_0 \Psi_2^0 + 4 \pi_0 \Psi_3^0- 4 \epsilon_0 \Psi_4^0,
 \nn\ea\ee
where the complex derivative is defined as
$ \up{c}\nabla:= \frac{\partial}{\partial x^2} +
i\frac{\partial}{\partial x^3}$, $ P(v,x^k):= \xi^{30}= -i
\xi^{40}$ and $ P\up{c}\nabla =\delta_0. $

 From \textbf{(d)} in p. 46 and complex conjugate of
\textbf{(e)} in p. 46 , we have
    \bea \dot \pi_0= \frac{d}{d v} (\a_0+\bar\b_0) = 2 \bar P \up{c}\nabla \epsilon_0 = -\bar\kappa_0.\eea
    Using \textbf{(c)} in p. 46, $\dot\pi_0 =-\kappa_0$. It implies $\kappa_0 -\bar\kappa_0
    =0$. Therefore $\kappa_0$ is real. Using \textbf{(a)} in p. 46 and the fact that $\Psi^0_3\neq 0$ for the rotating NEH , we get
 \bea \kappa_0
    =0 \text{, i.e.,}\;\;\;  \delta_0 \epsilon_0=0, \eea
 (i.e., $P\up{c}\nabla \epsilon_0 =0$) on rotating NEH.

\textbf{Surface gravity: from NEH to WIH}

Here we prove that the surface gravity for a rotating WIH is also
constant. We make a coordinate choice $r_0=- \frac{1}{\mu_0}$ on the NEH and
it becomes a WIH. This gives $\dot \mu_0 =0$. Applying time
derivative on \textbf{(h)} in p. 46 and using $\kappa_0=0$ from
\textbf{(b)} in p. 49, we then get $\dot\epsilon_0=0$.

It then gives us that $\epsilon_0$ is constant on WIH. So the
surface gravity $\kappa_{(\l)} \hat{=}\Re\epsilon_0$ is constant
on WIH. For NEH, the surface gravity is not necessary constant.

\subsection{Near a non-rotating dynamical
horizon: non-vacuum expansion}\label{AsyDH}


As before, we use advanced Eddington-Finkelstein type coordinates
$(v,r,\theta,\phi)$. To start with, we choose $n_a=\partial_a v$
along null hypersurface $v=const.$ We have $g^{ab} v_{,a} v_{,a}
=0$. This gives the gauge conditions $\nu=\mu-\bar\mu=
\gamma+\bar\gamma=\bar\a+\b -\bar\pi=0$. Further we can choose
$n^a$ flag plane parallel, this implies $\c=0$. The incoming
tetrad takes the form $ n^a= -\frac{d x^a}{d r}=g^{ab} v_{,b}=
\delta^\mu_2= (0,-1,0,0)$.
 $m, \bar m$ are chosen to be tangent to the two surface. Hence,
$\rho=\bar\rho, \pi=\bar \tau.$ Applying the null rotation type III
$\l\to A^{-1} \l, m\to e^{2i\theta} m$, we  can chose
$\theta$ to make $\Im [\epsilon] =0$.

\textbf{Gauge choices.} We conclude that our total gauge choices of
the frame in this section are
\bea \nu=\rho-\bar\rho=\mu-\bar\mu=\gamma=\epsilon-\bar\epsilon=0,
\pi=\bar\tau= \a+\bar\b. \label{Ch5-gaugeCD}\eea

In order to preserve orthonormal relation,
\be  \l^a n_a=1, m^a \bar m_a =-1, \l^a m_a= n^a m_a =0, \nn\ee
we can choose the tetrad as
\be\ba{l} \l^a=(1, U, X^3, X^4),\;
          n^a=(0,-1,0,0),\;
          m^a=(0,0, \xi^3,\xi^4). \nn\ea\ee
%


\textbf{Transfer to new coordinates $(v,r',\theta,\phi)$}

Since the dynamical horizon is not a null hypersurface, we have to
transfer to a 'good' coordinates which can capture this picture. We
defined a new radius coordinate by using
\bea r' = r- r_\Delta(v).\nn\eea
To tackle the problem of the power series expansion near a
non-null hypersurface, we make a coordinate transformation by
using $d r' = d r - \dot r_\Delta d v$ to transfer coordinates into
the new coordinates $(v, r', \theta,\phi)$. The tetrad in the new
coordinates $(v, r', \theta,\phi)$ are:
\be\ba{l} \l^a=(1, U - \dot{r}_\Delta, X^3, X^4),\\
          n^a=(0,-1,0,0),\;
          m^a=(0,0, \xi^3,\xi^4).  \label{tetradc}\nn\ea\ee
The metric has the form
\be g^{ab} =\l^a n^b +n^a \l^b -m^a\bar m^b - \bar m^a m^b,\nn\ee
where it's components are
\be\ba{l} g^{01}=-1, \; g^{00}=g^{0k} =0,\; g^{11} = -2 U+ 2 \dot r_\Delta,\\
 g^{1k} =- X^k, \;\; g^{mn} = -(\xi^m\bar\xi^n+ \bar\xi^m\xi^n). \nn\ea\ee
$H$ is a non-null hypersurface, since $g^{ab} {r'}_{,a} {r'}_{,b}
= -2 U + 2 \dot r_\Delta$ in this new coordinates.

\textbf{Gauge conditions of non-rotating dynamical horizons}

On the non-rotating dynamical horizons, the null normal $\l_a$ is
hypersurface orthogonal, therefore it is geodesic and twist-free.
Moreover, it is the null normal of the marginal trapping surface, therefore
it is expansion-free on the dynamical horizon. However, unlike the
isolated horizon, the shear is non-vanishing on the dynamical
horizon. Here we only consider a non-rotating dynamical horizon,
so $\pi$ vanishes on the boundary. In terms of NP, these
conditions are
\bea \kappa\hat{=} \rho\hat{=} \pi\hat{=}0\nn\eea
on the non-rotating DHs. Therefore, the fall off of NP coefficients and the
components of the null tetrad are
\be\ba{l}
          U, \xi^k, X^k, \mu,\epsilon,\a,\b, \lambda, \sigma, =O(1),\\
          \kappa, \rho, \pi =O(r').\nn
          \ea\ee

\textbf{The falloff of the Weyl tensor and stress energy tensor}

Because we want to compare with the Vaidya solution, we assume
that the falloff of the Weyl tensors  and Ricci tensors is the
same as that of Vaidya solution. Therefore, we take the Weyl
tensor as
\be\ba{l} \Psi_0=O(r'), \Psi_1=O(r'), \Psi_2=\Psi_3=\Psi_4 =O(1)
\label{DHWeyl} \ea\ee
and the Ricci tensors as
\bea \Phi_{00}=O(1),
\Phi_{22}=\Phi_{11}=\Phi_{02}=\Phi_{01}=\Phi_{21}=O(r').\label{DHRicci}\eea
We note that the coefficient  of order $O(1)$ of the tetrad
components $U, X^k$, i.e., $U_0, X^{0k}$ can be made to be
vanished by using coordinate transformation.

\textbf{The radial equations}

From \textbf{(n)},\textbf{(j},\textbf{(r)},\textbf{(o)},\textbf{(q)},\textbf{(p)},\textbf{(i)},\textbf{(f)},\textbf{(c)} in p. 46-47 :
\be\ba{lll}
\mu &=& \mu_0+ ({\mu_0}^2+\lambda_0\bar\lambda_0) r' + O({r'}^2), \\
\lambda&=&  \lambda_0+ (2\mu_0\lambda_0+ \Psi^0_4 ) r' + O(r'^2 ),\\
\a     &=& \a_0 +(\a_0\bar\mu_0-\lambda_0\bar\a_0 + \Psi^0_3) r'+ O(r'^2),\\
\b     &=&-\bar\a_0 + (\a_0\bar\lambda_0- \mu_0 \bar\a_0) r'+ O(r'^2),\\
\rho   &=& ( \sigma_0 \lambda_0 + \Psi^0_2 )r' + O(r'^2),\\
\sigma &=& \sigma_0 + \mu_0\sigma_0 r' +O(r'^2),\\
\pi   &=& \Psi^0_3 r'+ (\mu_0\pi_0+\lambda_0\bar\pi_0)r'^2+ O(r'^3),\\
\epsilon&=&\epsilon_0 + \Psi^0_2  r' + O(r'^2),\\
\kappa&=& \dot r_\Delta \bar\Psi^0_3 r' + O(r'^2), %
\ea\ee
From (305) and (303):
\be\ba{lll}
\xi^k&=& \xi^{k0} +(\bar\lambda_0\bar\xi^{k0}+\mu_0 \xi^{k0}) r' + O(r'^2),\\
U &=& 2 \epsilon_0 r' +  \Psi^0_2 r'^2+ O(r'^3),\\
X^k &=& (\Psi^0_3\xi^{k0} + \bar\Psi^0_3\xi^{k0})r'^2+ O(r'^3).
\ea\ee
From  \textbf{(e)}, \textbf{(f)},\textbf{(g)}, \textbf{(h)}, \textbf{(i)},\textbf{(j)} and \textbf{(k)} in p. 49-51 :
\be\ba{lll}
\Psi_0 &=& (-3\sigma_0 \Psi^0_2+ \bar\lambda_0 \Phi_{00}^0) r' + O(r'^2),\\
\Psi_1 &=& (-2\sigma_0\Psi^0_3 -\eth_0\Psi^0_2) r' + O(r'^2),\\
\Psi_2 &=& \Psi^0_2 + (-\eth_0\Psi^0_3 +3 \mu_0 \Psi^0_2 -\sigma_0\Psi^0_4 )r' + O(r'^2),\\
\Psi_3& =& \Psi^0_3 + (-\eth_0\Psi^0_4 +4 \mu_0 \Psi^0_3 )r' + O(r'^2),\\
\Psi_4 &= &\Psi^0_4 +\Psi^1_4 r' + O(r'^2), \\
\Phi_{00}&=& \Phi_{00}^0 + 2\mu_0 \Phi_{00}^0 r' + O(r'^2),\\
\Phi_{11}&=& \half \dot r_\Delta \Phi_{22}^0 r'^2 + O(r'^3),\\
\Phi_{01}&=& -\dot r_\Delta \Phi_{12}^0 r'^2 + O(r'^3),\\
\Phi_{22}&=& \Phi_{22}^0 r'^2+ O(r'^3), \Phi_{12}= \Phi_{12}^0
r'^2+ O(r'^3),\\
\Phi_{02} &=& \Phi_{02}^0 r'^2+ O(r'^2).
 \ea\ee

\textbf{The non-radial NP equations} 

From (304) and (306) in p. 45:
\be\ba{llll}
          P\up{c}\nabla \dot r_\Delta  =0, \;\;
          \dot P = \dot r_\Delta (\bar\lambda_0\bar P + \mu_0 P), \;\;
          \a_0 = \frac{1}{2} \bar P\up{\bar c}\nabla \ln P.
\ea\ee
From   \textbf{(a)},\textbf{(b)},\textbf{(c)}, \textbf{(g)},\textbf{(d)},\textbf{(e)},\textbf{(m)},\textbf{(l)},\textbf{(n)} in p. 46-47 :
\be\ba{lll}
\dot r_\Delta(\Psi^0_2+\sigma_0\lambda_0) = -\Phi_{00}^0 - \sigma_0\bar\sigma_0, \\ 
\dot \sigma_0 = 2\epsilon_0 \sigma_0 + \dot r_\Delta \mu_0\sigma_0,\\ 
\kappa_0 = - 2 P\up{c}\nabla \epsilon_0 =\dot r_\Delta \bar\Psi^0_3,\\ 
\dot \lambda_0  = -2\epsilon_0 \lambda_0 +\mu_0\bar\sigma_0+\dot
r_\Delta (2 \mu_0\lambda_0 + \Psi^0_4),   \\ 
\dot \a_0 =  \dot r_\Delta (\mu_0\a_0-\lambda_0\bar\a_0 + \half
\Psi^0_3) -\bar\a_0\bar\sigma_0,\\  
%
\Psi^0_3 =  \bar\eth_0\mu_0-\eth_0 \lambda_0,\\ 
\Psi^0_2 =
 \frac{P\bar P}{2} (2 \up{c}\nabla \ln \bar P \up{\bar c}\nabla \ln P-
\up{c}\nabla \ln \bar P - \up{\bar c} \nabla \ln P)
 -\lambda_0\sigma_0,\\ 
\Im \Psi^0_2 = -\Im(\lambda_0 \sigma_0), \\ 
\Psi^0_2 = \partial_v \mu_0  + 2\epsilon_0\mu_0 -\lambda_0\sigma_0
-\dot r_\Delta(\mu_0^2 +\lambda_0\bar\lambda_0).  
\nn \ea\ee
From  \textbf{(a)},\textbf{(b)},\textbf{(c)}, \textbf{(d)} in p. 49:
\be\ba{llll}
   \dot r_\Delta(2\sigma_0 \Psi_3^0 +\eth_0 \Psi_2^0) =- P\up{c}\nabla \Phi_{00}^0,\\ 
\partial_v \Psi^0_1 = \bar P \up{\bar c}\nabla \Psi^0_0+ 2\epsilon_0
\Psi^0_1 -3 \kappa_0 \Psi^0_2+\bar\pi_0 \Phi_{00}^0\\
\hspace{3em}-2\dot r_\Delta(\Phi_{01}^0-\Psi^1_1),\\ 
\partial_v \Psi^0_2 = \mu_0 \Phi_{00}^0 + \dot r_\Delta (-\eth_0 \Psi^0_3 + 3 \mu_0 \Psi^0_2-\sigma_0
\Psi^0_4),\\ 
\partial_v \Psi^0_3 = \bar\eth_0 \Psi^0_2 - 2 \epsilon_0\Psi^0_3 + \dot r_\Delta (-\eth_0 \Psi^0_4 +
4\mu_0\Psi^0_3),\\  
\partial_v \Psi^0_4 =  \bar\eth_0 \Psi^0_3 -3 \lambda_0 \Psi^0_2 - 4 \epsilon_0 \Psi^0_4 + \dot r_\Delta
\Psi^1_4. 
\ea\ee
For example, $\dot \sigma_0$ has a next order contribution from
asymptotic expansion since $\dot \sigma_0 = 2\epsilon_0 \sigma_0 +
\dot r_\Delta \sigma_1$, where $\sigma_1= \mu\sigma_0$.

\section{Constant spinors for quasi-local horizons} \label{conspinor}

\subsection{Constant spinors for the generic isolated horizons:
Frame alignment}

In this section, we adopt a similar idea of Bramson's asymptotic
frame alignment for null infinity \cite{Bramson75a} and apply it
to set up spinor frames on the quasi-local horizons. We define the
spinor frames
\bea Z_A\,^{\underline{A}} = (\lambda_A, \mu_A) \eea
where
$ \lambda_A = \lambda_1 o_A -\lambda_0 \iota_A$, $\mu_A = \mu_1
o_A - \mu_0 \iota_A $.
We expand $\lambda_1, \lambda_0$ as
\bea \lambda_1= \lambda_1^0(v,\theta,\phi) +
\lambda_1^1(v,\theta,\phi) r' + O(r'^2),\\
 \lambda_0=
\lambda_0^0(v,\theta,\phi) + \lambda_0^1(v,\theta,\phi) r' +
O(r'^2). \eea
Here $\lambda_1$ is type $(-1,0)$ and $\lambda_0$ is type $(1,0)$.

Firstly, we demand the conditions on spin frames to be
parallelly transported along null normal $\l^a$ direction on the
quasi-local horizons, so
\be\ba{l} \lim_{r'\to 0} D Z_A\,^{\underline{A}}=0, \label{SD}
\ea\ee
and also the conditions of the frames on different generators on
the quasi-local horizons are:
\be\ba{l} \lim_{r'\to 0} \delta Z_A\,^{\underline{A}}=\lim_{r'\to
0}\; \bar\delta Z_A\,^{\underline{A}}=0 . \label{SDelta}\ea\ee
%


We then get six conditions
\bea \dot{\lambda_0^0} -\epsilon_0 \lambda^0_0 &=&0,  \textrm{i.e., }\;\; \thorn_0 \lambda_0^0 =0, \label{Framee1}\\
     \dot{\lambda_1^0} +\epsilon_0 \lambda_1^0 &=&\pi_0 \lambda_0^0,  \textrm{i.e., }\;\; \thorn_0 \lambda_1^0 =\pi_0 \lambda_0^0,  \label{Framee2}\\
     \eth_0 \lambda_0^0             &=& 0,
     \label{Framee3}\\
     \eth_0 \lambda_1^0  - \mu_0 \lambda_0^0 &=& 0,       \label{Framee4}\\
     \eth'_0 \lambda_0^0      &=& 0,                  \label{Framee5}\\
     \eth'_0 \lambda_1^0  +\sigma'_0 \lambda_0^0 &=& 0.        \label{Framee6}
\eea
for the constant spinors $\lambda_A$ on the generic isolated
horizons.
In (\ref{Framee6}) and also the following, we
use the symbol $-\sigma'_0$ to represent the one of the NP coefficients, $\lambda_0$ (the leading order of NP
shear of $n$), in order to avoid confusion with the spinor $\lambda_A$


We use the condition (\ref{Framee1}) and the fact that
$\thorn_0\eth_0 =\eth_0\thorn_0$ on the horizon. Apply $\thorn_0$
on (\ref{Framee3}), we find
\bea 0= \thorn_0 \eth_0 \lambda_0^0 = \eth_0 \thorn_0 \lambda_0^0
=0.\eea
So condition (\ref{Framee1}) and (\ref{Framee3}) are compatible.

Apply $\thorn$ on (\ref{Framee4}) and use condition
(\ref{Framee1}) and (\ref{Framee2}), we have
\bea 0 &=& \Psi^0_2 \lambda_0^0.\eea
Hence the condition (\ref{Framee1}), (\ref{Framee2}) and
(\ref{Framee4}) are not compatible unless $\Psi_2^0=0$.


From the previous analysis, we conclude that \textit{the
compatible frame alignment conditions for the generic isolated
horizon} are (\ref{Framee1}), (\ref{Framee3}) and (\ref{Framee4}).
Equation (\ref{Framee3}) and (\ref{Framee4}) are Dougan-Mason's
holomorphic conditions \cite{Dougan&Mason}. Here we see that the
conditions of the spinor field to be asymptotically constant on
NEH implies the Dougan-Mason holomorphic conditions on the cuts of
the NEH. These equations will be used together with the
Nester-Witten two form to define the quasi-local energy-momentum.
The time related condition (\ref{Framee1}) will tell us how the
energy momentum changes with time along NEH and will be useful to
calculate the energy flux across the horizon.

%

\subsection{Constant spinors for a non-rotating dynamical horizon}

The dynamical horizon is a space-like or time-like hypersurface if
the horizon is expanding $\dot r_\Delta>0$ or contracting. From
this we know that it's generator is non-null and it is a time-like
or space-like generator respectively. To understand this, we have
to work in the $(v, r',\theta,\phi)$ coordinate. The dynamical
horizon generator $R$ is tangent to the dynamical horizon where
%
%
$\L_R v =1$ so $R=\frac{d}{dv}$. We have the relation
 \bea  R^a = \l^a- \dot r_\Delta n^a =\frac{\partial}{\partial
 v},\;\;\nn\\
       R_a = \l_a-\dot r_\Delta n_a = -2 \dot r_\Delta d v - d r',
       R^a R_a = - 2 \dot r_\Delta.\nn\eea
So if $\dot r_\Delta>0$ then the $H$ is a space-like hypersurface.
Since $g^{ab} r'_{,a} r'_{,b} = 2 \dot r_\Delta$.

We define the spin frame $Z_A\,^{\underline{A}} = (\lambda_A,
\mu_A)$ where
\bea \lambda_A &=& \lambda_1 o_A - \lambda_0\iota_A,\\
 \mu_A &=& \mu_1 o_A -\mu_0 \iota_A,\eea
and
\bea \lambda_0 &=& \lambda_0^0(v,\theta,\phi)
+\lambda_0^1(v,\theta,\phi) r' + O(r'^2),
\\
     \lambda_1 &=& \lambda_1^0(v,\theta,\phi) + \lambda_1^1(v,\theta,\phi) r' +
O(r'^2),\eea
where $\lambda_1$ is type $(-1,0)$ and $\lambda_0$ is type
$(1,0)$.

Firstly, we demand that the conditions on spin frames to be
parallelly transported along the non-rotating dynamical horizon
generators $R^a$ direction on $H$, so
\be\ba{l} \lim_{r'\to 0} R Z_A\,^{\underline{A}}= \lim_{r'\to 0}
(D-\dot r_\Delta \Delta) Z_A\,^{\underline{A}}=0, \label{TCDH}
\ea\ee
where we use $R^a = (\l^a - \dot r_\Delta n^a)$. The conditions of
the frames on different generators on $H$ are:
\be\ba{l} \lim_{r'\to 0} \delta Z_A\,^{\underline{A}}= \lim_{r'\to
0}\; \bar\delta Z_A\,^{\underline{A}}=0 . \label{SCDH}\ea\ee
%

Therefore, from $o^A (D-\dot r_\Delta \Delta) \lambda_A$ in
equation (\ref{TCDH}) we get
\bea  (D-\dot r_\Delta \Delta -\epsilon + \dot r_\Delta \c)
\lambda_0 + (\kappa- \dot r_\Delta \tau) \lambda_1 =0\eea
where  we use $\c = 0, \pi=\bar\tau\hat{=}0,\kappa=O(r')$. The
total six equations that include two time-related conditions and
four spatial-related conditions are
\be\ba{llll} \dot{\lambda_0^0} -\epsilon_0 \lambda^0_0  &=&0,
  \textrm{i.e., }\;\; \thorn \lambda_0^0 =\dot r_\Delta \thorn' \lambda_0^0=0, & (a) \nonumber\\
     \dot{\lambda_1^0} +\epsilon_0 \lambda_1^0 &=&0,
     \textrm{i.e., }\;\; \thorn \lambda_1^0 =\dot r_\Delta \thorn' \lambda_1^0=0,  & (b)\nonumber\\
     \eth_0 \lambda_0^0  + \sigma_0 \lambda_1^0   &=& 0   \;\;\textrm{from $\eth \lambda_0 +
     \sigma\lambda_1=0$},
              & (c)\nonumber \\
     \eth_0 \lambda_1^0  - \mu_0 \lambda_0^0 &=& 0  \;\;\textrm{from $\eth \lambda_1
     -\mu\lambda_0=0$},
       & (d)\nonumber \\
     \eth'_0 \lambda_0^0      &=& 0        \;\;\textrm{from $\bar\eth \lambda_0 +
     \rho\lambda_1=0$},
       & (e)\nonumber\\
     \eth'_0 \lambda_1^0  - \lambda_0 \lambda_0^0 &=& 0  \;\;\textrm{from $\bar\eth \lambda_1 +
     \lambda\lambda_0=0$},
         & (f)\nonumber
\ea\ee


From the commutation relations, we have $\thorn_0\eth_0 -
\eth_0\thorn_0 \hat{=}\sigma_0 \eth_0'$. From $(a)$, we have
$\eth_0\thorn_0 \lambda_0^0 =0$. From $(c)$ and use the fact
$\thorn_0\sigma_0=0$, we have $ \thorn_0\eth_0 \lambda_0^0 = -
\sigma_0 \thorn_0 \lambda^0_1$.
To make this relation compatible with commutation relation, we
need an extra condition
\bea \thorn_0\lambda_1^0 = - \bar\eth_0\lambda^0_0. & (g) \nn \eea
To examine whether this extra condition is compatible with
Dougan-Mason holomorphic condition, we apply $\thorn_0$ on $(d)$.
From $(d)$, we have
\bea \thorn_0 \eth_0 \lambda^0_1 = (\sigma_0\lambda_0 + \Psi^0_2)
\lambda^0_0.\eea
From $(g)$ and together use the fact $\bar\eth_0\sigma_0=0$ and
commutation relation of $[\eth,\eth']$, we have
\bea \eth_0 \thorn_0 \lambda^0_1 = \sigma_0 \eth_0'\lambda_1^0+
(\sigma_0\lambda_0 + \Psi^0_2) \lambda^0_0.\eea
Therefore, this extra condition is compatible with Dougan-Mason
holomorphic condition.


From the previous analysis, we concludes that the compatible frame
alignment conditions for the non-rotating dynamical horizon would
be one time related condition (\ref{TCDdhs}) (i.e., $(a)$), two
Dougan-Mason holomorphic conditions (\ref{DM1dhs}) (i.e., $(c)$)
and (\ref{DM2dhs}) (i.e., $(d)$) and one extra condition
(\ref{extradhs}) (i.e., $(g)$):
\bea \thorn_0 \lambda^0_0 &=& \dot r_\Delta \thorn' \lambda_0^0
=0,
\textrm{ i.e.,}\;\; \dot{\lambda}^0_0 -\epsilon_0 \lambda^0_0=0,\label{TCDdhs}\\
\eth_0 \lambda^0_0 + \sigma_0\lambda_1^0&=&0,\label{DM1dhs}\\
\eth_0 \lambda^0_1 - \mu_0 \lambda^0_0&=& 0,\label{DM2dhs}\\
\thorn_0 \lambda^0_1 &=& - \bar\eth_0\lambda^0_0, \label{extradhs}
 \eea
where first three conditions are similar with the compatible
constant spinor conditions for the generic isolated horizon.
If the shear term vanishes, we do not need this extra
condition.

\textbf{Remark.} The Dougan-Mason holomorphic conditions
(\ref{DM1dhs}) and (\ref{DM2dhs}) will tell us how to gauge fix
the horizon quasi-local energy expression and the time related
condition (\ref{TCDdhs}) will tell us how the energy momentum
change with time along the dynamical horizon.

\section{Energy-momentum and flux} \label{energy-flux}

\subsection{The quasi-local energy-momentum of an isolated
horizon}\label{emIH}

By using Nester-Witten two form together with the compatible
constant spinor conditions which are Dougan-Mason's holomorphic
conditions (\ref{Framee3}) and (\ref{Framee4}) for the NEH, the
quasi-local momentum integral near a NEH is
\be\ba{lll} I(r') &=& -\frac{1}{8\pi} \oint_{S_r}
[\lambda_{0'}\eth \lambda_1 - \lambda_{1}\eth \lambda_{0'} +
\lambda_{0}\eth' \lambda_{1'} - \lambda_{1'}\eth' \lambda_{0}\\
 & &- \lambda_0\lambda_{0'} (\mu+\bar \mu) - \lambda_1\lambda_{1'}
(\rho+\bar\rho)] d S \\
&=&  \frac{1}{4\pi} \oint_S [-\mu_0 \lambda_0^0 \bar\lambda^0_{0'}
+ O(r')]d S. \label{mom-int} \ea\ee
Moreover, the horizon momentum  $P_{\underline{AA'}}$ can be
written as
\bea P_{\underline{AA'}} (S_\Delta) =
I(r_\Delta)\lambda_{\underline{A}}\bar\lambda_{\underline{A'}}
\eea
where $\lambda_{\underline{A}}$ is constant spinor on two surface
of NEH. From the result of the asymptotic expansion for the
generic isolated horizons, we can re-interpret the
\textit{quasi-local energy-momentum integral of the generic
isolated horizons (NEH)} as
\bea I(r_\Delta) = -\frac{1}{4\pi} \oint_S \frac{\Psi^0_2
-\dot\mu_0 + \eth_0 \pi_0 + \pi_0\bar\pi_0}{2 \epsilon_0}
\lambda_0^0 \bar\lambda^0_{0'}
 d S_\Delta \label{quasilocal-RNEH}\eea
where $\Psi_2^0 = M + i L$ and $\eth_0\pi_0 = A -i L$ with $M, L,
A$ are function of $(v,\theta,\phi)$. We compare with Kerr solution in the Appendix \ref{Kerr}.

\subsection{News function of the generic
isolated horizons} \label{NEHflux}

In order to match the Kerr solution that its flux vanishes, we
rescale the spinor field. Firstly, the constant spinors
$\lambda^0_0$ and $\lambda^0_1$ are rescaled by using the
following relation
\be\ba{lllll} \tilde{\lambda}^0_0 &=& \lambda^0_0 e^{-\int
\epsilon_0 d v},\;\;
      \tilde{\lambda}^0_{1} &=& \lambda^0_{1} e^{-\int \epsilon_0 d
      v}, 
      \ea\ee
and it yields the new rescaled momentum integral
\bea \tilde{I}(r_\Delta) &=& e^{-2 \int \epsilon_0 d v}
I(r_\Delta)
           = -\frac{1}{4\pi} \oint \mu_0 \tilde{\lambda}^0_0
\tilde{\bar\lambda}^0_{0'} d S_\Delta. \eea
The three compatible conditions (\ref{Framee1}), (\ref{Framee3})
and (\ref{Framee4}) then become
\bea \dot{\tilde{\lambda}}^0_0 =0,\;\; \eth_0
\tilde{\lambda}^0_0=0, \;\; \eth_0 \tilde{\lambda}^0_1 - \mu_0
\tilde{\lambda}^0_0= 0  \eea
where we use $\delta_0 \epsilon_0=0$ from asymptotic expansion and
they are still compatible under rescaling.

By using this new rescaling constant spinor frame, we apply the
time derivative on the quasi-local energy-momentum of NEH
(\ref{mom-int}) and thus we get
\bea \dot{\tilde{I}}(r_\Delta) = - \frac{1}{4\pi}\oint \dot\mu_0
\tilde{\lambda}^0_0 \tilde{\bar\lambda}^0_{0'} d S_\Delta
\label{NewdotI}\eea
We said that (\ref{NewdotI}) is \textit{quasi-local energy flux
near NEH}.  Here $\dot \mu_0$ is related with the mass loss or
gain, hence it is the \textit{news function of NEH}. However, since the black hole laws do not
hold on NEH, we conclude that  \textit{the news function of NEH is not physically
reasonable. This is due to that one cannot measure the correct
temperature of a black hole since he or she uses a bad
thermometer, which corresponds to the normalization of $\l$. If one choose the canonical $[\l]$ which yield WIH, then $\dot\mu_0$ vanishes.}

\subsection{Conserved quantities of the generic isolated
horizons}\label{conservedQ}

We now consider the absolute conservation law that $\dot G_m=0$ on
the generic isolated horizons. From \textbf{(b)} in p. 49, we have
$\dot\Psi_2^0=0$, therefore, we can find ten conserve quantities
which are
\be G_m= \int {_2} Y_{2,m} \Psi_2^0 d S .  \;\;\;\;\;
(m=-2,-1....,2)\ee
Here these conserved quantities corresponds to three different
type of generic IHs. These conserved quantities
are related with mass and angular momentum of the generic IHs.


\subsection{Quasi-local energy momentum of a non-rotating dynamical horizon}

By using the Nester-Witten two form and Dougan-Mason's holomorphic
conditions (\ref{DM1dhs}) and (\ref{DM2dhs}) which we found in
pervious section, the quasi-local energy-momentum integral on
horizon can be expressed as
\bea I(r_\Delta) = - \frac{1}{4\pi}\oint \mu_0 \lambda_{0}^0
\bar\lambda^0_{0'} d S_\Delta. \eea
Then use $\textbf{(h)}$ in p. 46,
the quasi-local
energy-momentum integral for a non-rotating dynamical horizon can
be expressed as
\bea I(r_\Delta) &=& - \frac{1}{4\pi}\oint \frac{1}{2
\epsilon_0}[\Psi_2^0 -\dot\mu_0 -\sigma_0{\sigma'}_0\\
&& + \dot r_\Delta(\mu_0^2+\sigma'{_0}\bar\sigma'{_0})]
\lambda_{0}^0 \bar\lambda^0_{0'} d S_\Delta.
\label{ch5-quasi-local}\eea
%

\subsection{Flux expression of a non-rotating dynamical horizon}

\textbf{Analysis and properties of NP equations}

\begin{description}
    \item[<1>]  From \textbf{(a)} in p. 46, because of dominate energy condition,
we have $\dot r_\Delta (\Psi^0_2 - \sigma_0\sigma'{_0}) =
-\Phi^0_{00} - \sigma_0\bar\sigma_0 \leq 0$ and $\dot r_\Delta\geq
0$. It implies $\Psi_2^0  - \sigma_0\sigma'{_0}\leq 0$.

    \item[<2>] From imaginary of \textbf{(l)} in p. 46, it implies $Im \Psi_2^0 =
    (\sigma_0\sigma'{_0}
 -\bar\sigma_0\bar\sigma'{_0})$. If $\Im \Psi_2^0 =0$ for shear non-zero, then it implies $\sigma'{_0}
 =-  A\bar\sigma_0$ where $A$ is real and time independent.

  \item[<3>] Use \textbf{<2>} and from the result of asymptotic expansion
  $\textbf{(b)}$ in p. 46 and $\textbf{(g)}$ in p. 46, we get $0=(-4 A \epsilon_0 + A \dot r_\Delta \mu_0 + \mu_0)\sigma_0 + \dot r_\Delta
 \bar\Psi_4^0$. (1) If  $\dot
r_\Delta =0$, it implies $(-4A\epsilon_0 +\mu_0)\sigma_0=0$. So
either $\sigma_0=0$ or $\mu_0=4A\epsilon_0$. For the later case
$\dot r_\Delta =0$ does not imply shear vanishes. Since $\dot r_\Delta$ should imply shear vanishing, this case does not satisfy the boundary conditions of  NEHs. (2) If
$\sigma_0=0$, then there are two situation. (a) $\dot r_\Delta=0$
then it goes back to NEH. (b) $\Psi_4^0=0$ but $\dot r_\Delta$ not
vanish. It is similar to  Vaidya.

    \item[<4>]  From \textbf{(g)} in p. 46, if $\sigma'{_0} =0$, i.e., $A=0$, then
$\mu_0\bar\sigma_0 =- \dot r_\Delta \Psi_4^0$. (1) If $\dot
r_\Delta =0$ since $\mu_0$ cannot be zero then $\sigma_0=0$. (2)
If $\sigma_0=0$, then there are two situation. (2') $\dot
r_\Delta=0$ then it goes back to NEH. (2'') $\Psi_4^0=0$ but $\dot
r_\Delta$ not vanish. It is similar to Vaidya.

From \textbf{<3>} and \textbf{<4>}, it would be more reasonable for us to chose $\sigma'{_0} =0$,
i.e., $A=0$.

   \item[<5>] If we choose the coordinate $r_\Delta =-\frac{1}{\mu_0}$, then
   $\mu_0$ is not related with $\theta$ or $\phi$. So $\bar\eth_0 \mu_0
   =0$. Together with the choice $\sigma'{_0} =0$, we then have $\Psi_3^0=0$
   from \textbf{(m)} in p. 47.

\end{description}


From the above analysis and reasons, we choose $r_\Delta =-\frac{1}{\mu_0}$ and set
\bea \sigma'{_0}=0\nn\eea
to derive flux expression. From \textbf{(g)} in p. 46,
\textbf{(b)} in p. 49 and \textbf{(h)} in p. 46, we have
\bea \mu_0 \bar\sigma_0 &=& -\dot r_\Delta \Psi_4^0,\\
     \dot \Psi_2^0 &=& \mu_0 \Phi_{00}^0 + 6 \mu_0^2\epsilon_0\dot r_\Delta + \mu_0\sigma_0\bar\sigma_0,\\
     \Psi_2^0 &=& 2 \epsilon_0\mu_0. \eea
%
%
%
The energy-momentum integral (\ref{ch5-quasi-local}) then becomes
\bea I(r_\Delta)&=& - \frac{1}{4\pi}\oint \mu_0 \lambda_{0}^0
\bar\lambda_{0'}^0 d S_\Delta \;\;\;\;\;\textrm{(i})\nn\\
&=& - \frac{1}{4\pi}\oint \frac{\Psi^0_2}{2\epsilon_0}
\lambda_{0}^0 \bar\lambda_{0'}^0 d S_\Delta
\;\;\;\;\;\textrm{(ii)} \nn\eea
Now we apply time derivative on the energy-momentum integral and
also we use the time related conditions (\ref{TCDdhs}) of spinor
field.

Apply time derivative on (i), we get
\bea \dot I(r_\Delta) =  \frac{1}{4\pi} \oint \dot \mu_0
\lambda_{0}^0 \bar\lambda_{0'}^0 d S_\Delta. \label{flux-i}\eea

Apply time derivative on (ii), we get
 \bea
\dot I(r_\Delta) = -  \frac{1}{4\pi}\int \frac{\mu_0}{2
\epsilon_0}[ \sigma_0\bar\sigma_0 + \Phi_{00}^0 ]
 \lambda_{0}^0
\bar\lambda_{0'}^0 d S_\Delta. \label{flux-non-rotate}\eea
%

Finally, we get the quasi-local energy-momentum flux formula which
is obtained from asymptotic expansion for the non-rotating
non-spherical symmetric dynamical horizon (shear non-vanishing on
the boundary) is
 \bea \dot I(r_\Delta) &=& \frac{1}{4\pi} \oint \dot \mu_0 \lambda_{0}^0 \bar\lambda_{0'}^0
d S_\Delta \nn\\
 &=&   -\frac{1}{4\pi}\int \frac{\mu_0}{2\epsilon_0}
[\sigma_0\bar\sigma_0+\Phi_{00}^0] \lambda_{0}^0
\bar\lambda_{0'}^0 d S_\Delta \label{flux} \eea
which is quite similar with Ashtekar's  flux equation for
dynamical horizon.

 Therefore, $\dot \mu_0$ is \textit{the news function of non-rotating DHs} that determines
gravitational radiation and matter field radiation. The shear
square term $|\sigma_0|^2$ is believed to be related with
gravitational radiation, since the gravitational
radiation comes from the non-spherical symmetric gravitational
collapse. In the later section, we will see more detail about how
this equation can be compared with Ashtekar's flux formula.

\subsection{The relationship between Ashtekar-Krishnan flux
and our flux formula}\label{Ash&mine}

\textbf{From (i)}

From the choice of $\mu_0 =-\frac{1}{r_\Delta}$, we have
$\dot\mu_0 =\frac{\dot r_\Delta}{r_\Delta^2}=  \frac{\dot
r_\Delta}{2} \;^{(2)} R $ where the two scalar curvature is
$^{(2)} R = \frac{2}{r_\Delta^2}$ (The metric of a two sphere with
radius $r_\Delta$ is $d l^2 = - r_\Delta^2 (d\theta^2 +
\sin^2\theta d\phi^2)$.). We use these relations and substitute them
into the the time derivative of (i) (recall (\ref{flux-i})) in
previous section, we then obtain
\bea \dot I(r_\Delta) = \frac{d I}{d v}&=&  \frac{1}{4\pi} \oint
\dot
\mu_0 \lambda_{0}^0 \bar\lambda_{0'}^0 d S_\Delta \nn\\
&=& \frac{1}{8\pi} \oint  \frac{d r_\Delta}{d v} \;^{(2)} R
\lambda_{0}^0 \bar\lambda_{0'}^0 d S_\Delta. \nn
 \eea
Integrate the above equation with respect to $v$, we then have
\bea d I =\frac{1}{8\pi} \int \;^{(2)}R \lambda_{0}^0
\bar\lambda_{0'}^0 d S_\Delta d r_\Delta. \label{dI-i}\eea

We recall that the Ashtekar's total flux formula \cite{Ashtekar02}
is
 \bea F_{matter} + F_{grav} = \frac{1}{16\pi}\int_{\Delta H}  \,^{(2)}
 R N
d^3 V \label{Ash-i}\eea
which $F_{matter} + F_{grav}$ is equal to flux $d I$ and $d^3 V =
d r_\Delta d S$ on horizon. Therefore, if $N=2 \lambda_{0}^0
\bar\lambda_{0'}^0$, the our flux formula from equation (i) is
completely the same with Ashtekar-Krishnan's formula
(\ref{Ash-i}).


\textbf{From (ii)}

Recall the Ashtekar-Krishnan gravitational flux expression
(\ref{ch3-grav-flux}) is
\bea F_{grav} = \frac{1}{4\pi}\int_{\Delta H} N (|\sigma|^2
+|\pi|^2) d^3 V \eea
where it satisfies the gauge condition of our asymptotic
expansion. The matter flux expression of Vaidya solution is
\bea F_{\textrm{matter}} :&=& \int_H T_{ab} T^a \l^b N d^3 V\\
               &=& \frac{1}{4\pi} \int \Phi_{00} N d^3 V \eea
where we use $4 \pi T_{ab} \l^a \l^b = \Phi_{00}^0$. The total
Ashtekar-Krishnan flux of non-rotating DH  \cite{Ashtekar02}
\cite{WuWang-RDH} then becomes
\bea F_{\textrm{total}} = \frac{1}{4\pi} \int [ |\sigma|^2 +
\Phi_{00}] N d^3 V.\label{Ash-ii} \eea
Recall our formula (\ref{flux-non-rotate}) in previous
section and  then integrate it with respect to $v$, we have
 \bea
d I(r_\Delta) &=& -  \frac{1}{4\pi}\int \frac{\mu_0}{2
\epsilon_0}[ \sigma_0\bar\sigma_0 + \Phi_{00}^0 ]
 \lambda_{0}^0
\bar\lambda_{0'}^0 d S_\Delta d v \nn\\
&=& -  \frac{1}{4\pi}\int \frac{\mu_0}{2 \epsilon_0 \dot
r_\Delta}[ \sigma_0\bar\sigma_0 + \Phi_{00}^0 ]
 \lambda_{0}^0
\bar\lambda_{0'}^0 d S_\Delta d r_\Delta \nn\eea
where $d v = \frac{d r_\Delta}{\dot r_\Delta}$ (Since $d r'$
vanishes on horizon.).

\begin{enumerate}
    \item
Using \textbf{(h)} in p. 46 and \textbf{(l)} in p. 46, we have
 \bea 2\epsilon_0\mu_0=-P\up{c}\nabla\a_0 -\bar P \up{\bar c}\nabla\bar\a_0 +
  4\a_0\bar\a_0 = -\frac{1}{2 r_\Delta^2}=-\half\mu_0^2\nn\eea
where we use the fact that for a sphere metric $\a_0=
-\frac{\sqrt{2} \cot\theta}{4 r_\Delta}$. It then implies
\bea \mu_0=-4\epsilon_0.\eea

    \item Because $v$ is arbitrary, one can always rescale $v$ so that we can chose $\dot
r_\Delta=1$.

\end{enumerate}

From 1. and 2. and with the choice of $N=2 \lambda_{0}^0
\bar\lambda_{0'}^0 > 0$, we then have $-\frac{\mu_0}{2\epsilon_0
\dot r_\Delta}=2$. Our formula is then the same as
Ashtekar-Krishnan's formula (\ref{Ash-ii}). Therefore, the surface
gravity $\kappa_{(\l)}$ is $\kappa_{(\l)}=\frac{1}{2 r_\Delta}$.
The differential of the horizon area is $d A = 8\pi r_\Delta d
r_\Delta$ and $\kappa_{(\l)} d A = 4\pi d r_\Delta$. For the time
evolute vector $t^a = N \l^a$, the difference of the horizon
energy $E^t$ can be expressed as
\bea \int d E^t &=& \int d I(r_\Delta) = \frac{1}{4\pi} \int [
|\sigma|^2 +
\Phi_{00}] N d^3 V\\
&=&\int \half d r_\Delta = \int \frac{\kappa_{(\l)}}{8\pi} d A
\eea
which it is always positive. We can get \textit{a generalized
black hole first law for non-rotating DHs}
\bea \frac{\kappa_{(\l)}}{8\pi} d A = d E^t. \label{firstlaw}\eea

\section{Conclusions}

We know that the news function determines the gravitational
radiation near the null infinity. It is the time derivative of the
shear of outgoing null tetrad ($\dot{\bar\sigma}_0$). Similarly,
the horizon news function can determine the gravitational
radiation and energy loss or gain near the horizon. Our work of
the asymptotic expansions near quasi-local horizons does not
require the space-time is asymptotically flat. Using the method of
asymptotic expansions we find that the corresponding news
functions near quasi-local horizons are the time derivative of the
expansion of the incoming null normal ($\dot\mu_0$). If we apply
the time derivative on the Bondi mass, it will give us a minus
quadratic of time derivative of the shear term. It is always
\textit{negative}! Therefore, for an isolated gravitating system,
it will always lost mass and the gravitational wave will carry
gravitational radiation out to infinity.

The news function of dynamical horizon determines the quasi-local
energy flux cross the dynamical horizon to be always \textit{positive}, if the dominate energy condition holds. Then the
dynamical horizon will gain mass and the horizon will grow.
In this paper, we have investigated the quasi-local
energy of quasi-local horizons for which the physical energy
conditions hold.

Asymptotically constant spinors can be used to define the
quasi-local energy-momentum of the horizon. Searching for the
compatible conditions of constant spinors of horizons offers us a
way to chose for the proper reference frames when measuring these
quasi-local quantities. In practical, Dougan-Mason's holomorphic
conditions refer to how to fix the gauge (good measurement) in
quasi-local energy-momentum expression (Nester-Witten two form) on
each cross section of horizon. The time related condition can tell
us how the quasi-local energy-momentum change with time along
horizon.

We find that the news function exists for the NEHs. It indicates two possibilities: one is that there are gravitational
radiations outside the NEHs, and the other is that gravitational radiations cross to the equilibrium black hole but the horizon area will not increase. For the first one, we \textit{cannot} prove
whether news function corresponding to gravitational radiations outside the NEHs. For
the second one, it would be against black hole thermodynamic laws. However, since the news function of the NEHs can be made to be vanished
while we make a special choice of affine parameter $r_\Delta=
-\frac{1}{\mu_0}$, we conclude that  \textit{this
is due to a bad measurement (bad gauge choice) of gravitational
flux}. This result refers to that a generic NEH admits
a unique $[\l]$ such that $(\Delta,[\l])$ is a WIH on which the
incoming expansion $\mu_0$ is time independent \cite{Ashtekar99b}.

\textbf{Laws of black hole dynamics}

    \textbf{Zeroth law} For NEHs or DHs, the zeroth law will not hold. If we make a coordinate choice
    $r_\Delta =- 1/\mu_0$ for NEH, then WIH will preserve the zeroth law mechanics.

    \textbf{First law} From our construction, we can derive the generalized first law eq.
    (\ref{firstlaw}) for non-rotating DHs.

    \textbf{Second law} It is obvious that our flux formula for a dynamical horizon is related with the black hole area
law. If the news function $\dot \mu_0$ is positive, i.e., the
positive energy flux. From the fact of the area of cross section
changing with time $\frac{d}{d v} d S_\Delta = 2 r_\Delta
\dot{\mu_0} d S_\Delta$, we know that the black hole area is
always increasing if the dominate energy condition holds.
Therefore, the black hole second law can extend to a dynamical
horizon.

For a spherical symmetric dynamical horizon, the contribution of
energy radiation is purely from the matter field. Therefore the
gravitational contribution is from the non-spherical symmetric
term: the shear term. Our result about the gravitational radiation
of non-rotating DHs that comes from the next order contribution of
asymptotic expansion is proportional to the shear square and
always positive. This result agrees with perturbation method of
energy flux cross event horizon [Done by Hawking and Hartle, see
Chandrasekhar \cite{Chandrasekhar}] and Ashtekar-Krishnan's flux
formula for non-rotating DH.

One can refer to Chapter 6 of \cite{Wu2007} or ref \cite{WuWang-RDH}
for the relations between energy fluxes and area balance of rotating
dynamical horizons. The future work on gravitational radiation and angular momentum flux of
rotating DHs is in progress \cite{WuWang-RDH}.

\acknowledgments

We would like to thank anonymous referee to point out some
physical consequence of this work. YHW is partially supported by
NSC under grant no. NSC 097-2112-M-008-001. CHW is supported by
NSC under grant no. NSC 097-2811-M-008-015.


\begin{appendix}

\section{Quasi-local  mass for the Kerr solution}\label{Kerr}

Here we present three different calculations for Kerr horizon
mass. In Subsection \ref{bergq}, we re-calculate Bergqvist's
calculation for Kerr horizon mass \cite{Bergqvist92}. He used the
Dougan-Mason mass definition and solve the holomorphic conditions
to get the exact expressions for the constant spinors. To
calculate the Dougan-Mason mass on the bounded two sphere, i.e.,
at the quasi-local level, we need a tetrad to satisfy the gauge
conditions ($\rho,\mu$ are real). Therefore, we need to make a
null rotation of Kerr solution in advance Eddington-Frankelstein
coordinate. By using the constant spinors, the new NP coefficients and
the Dougan-Mason mass, we can calculate the quasi-local mass for
Kerr. In order to compare with our work, we expand it with repect to 
angular parameter $a$ when $a$ is small.

To make our tetrad to be compatible with the method of the
asymptotic expansion for Kerr, we transfer it to the Bondi
coordinate. Thus we have the desire gauge conditions in Bondi
coordinate when $a$ is small in Kerr metric. By using the results
of the Kerr solution in the approximate Bondi frame in Subsection \ref{KerrBondi}, 
we can solve the holomorphic
conditions and calculate the quasi-local mass by using
Dougan-Mason's definition in Subsection \ref{DMKerr-bondi}.

In Subsection \ref{myquasi-localKerr}, we check the Kerr solution by using our quasi-local
mass that obtain from asymptotic expansion (\ref{quasilocal-RNEH})
and see whether it can compare with the other two resuls in the previous sections. Our
conclusion is that all these three calculations are the same up to
the second order when power expanding with respect to the angular
momentum parameter $a$.

\subsection{Kerr solution in the approximate Bondi
frame} \label{KerrBondi}

%
We transfer Kerr solution from advanced Eddinton-Frankelstein cooordinates  to Bondi coordinate near horizon (detail see Chapter 6 of  \cite{Wu2007}). The Kerr metric in Bondi coordinate $(\tilde{v},
\tilde{r'},\tilde{\theta},\tilde{\phi})$ near the horizon is therefore
given by
\bea g^{00}&=& O(r'^2),\;\; g^{03}= O(r'^2),\;\;g^{12}= O(r'^2),\;\;g^{02}= O(r'^2),\nn\\
     g^{01}&=& - \frac{r_\Delta\,^2 + a^2}{\Sigma_\Delta} +[ \frac{2 r_\Delta a^2\sin^2\theta}{\Sigma_\Delta^2} +
     \frac{a^2\sin^2\theta(r_\Delta- M)}{(r_\Delta^2 +a^2)\Sigma_\Delta}]r'\nn\\
&+& O(r'^2),\nn\\
     g^{13}&=& -\frac{a}{\Sigma_\Delta} + [\frac{a(r_\Delta - M )(2r_\Delta\,^2 + 2 a^2
     - a^2\sin^2 \theta)}{(r_\Delta\,^2 +a^2)^2 \Sigma_\Delta} \nn\\
      &+&  \frac{2 a r_\Delta}{\Sigma_\Delta\,^2}]r' + O(r'^2),\nn\\
     g^{22}&=& -\frac{1}{\Sigma_\Delta} + 2 \frac{r_\Delta}{\Sigma_\Delta\,^2} r' + O(r'^2),\nn\\
     g^{33}&=& -\frac{1}{\sin^2\theta \Sigma_\Delta} + \frac{a^2 (2r_\Delta\,^2 + 2 a^2 -
      a^2\sin^2 \theta)}{2(r_\Delta\,^2 +a^2)^2 \Sigma_\Delta} + O(r'),\nn\\
     g^{23}&=& - \frac{2 a^3 \sin\theta\cos\theta}{(r_\Delta\,^2 +a^2)^2 \Sigma_\Delta} r' + O(r'^2),\nn\\
     g^{11}&=& -2 \frac{r_\Delta- M }{\Sigma_\Delta} r' + O(r'^2), \nn
     \eea
where $r_\Delta:= r_+ = M+\sqrt{M^2 -a^2}$ and $\Sigma_\Delta:=
r_\Delta^2 + a^2\cos^2\theta$.

 Now we consider the case of slow rotation so that $a$ is
small, the tetrad components in
the Bondi coordinate $(\tilde{v},
\tilde{r'},\tilde{\theta},\tilde{\phi})$ are:
\bea \l^a &=& (1, U r', 0, \frac{a}{r_\Delta\, ^2} + D r') ,\\
     n^a &=& (0, -1, 0, 0) ,\\
     m^a &=& \frac{1}{\sqrt{2} \eta_\Delta}(0,0, 1- \frac{r'}{\eta_\Delta}, \frac{- i}{\sin\theta}(1- \frac{r'}{\eta_\Delta})), \eea
where $\eta_\Delta:= r_\Delta + i a \cos\theta$,
$U:=\frac{r_\Delta - M}{r_\Delta^2}$ and $D := \frac{a(2 r_\Delta
- M)}{r_\Delta^4}$. The NP coefficients and Weyl tensors are:
\bea \kappa&=&\sigma=\lambda=\nu\hat{=}0,\nn\\
 \rho&=&\frac{U(-r_\Delta+ r') r'}{(\eta_\Delta -r')(\bar\eta_\Delta -r')} \hat{=} 0,\nn\\
 \mu &=& \frac{-r_\Delta + r'}{(\eta_\Delta -r')(\bar\eta_\Delta -r')} \hat{=} -\frac{r_\Delta}{\Sigma_\Delta},\nn\\
 \pi&=&\bar\tau = \frac{i \sqrt{2} D \eta_\Delta^2 \sin\theta}{4 (\eta_\Delta -r')} \hat{=} \frac{i \sqrt{2}
 D \eta_\Delta \sin\theta }{4},\nn\\
\b &\hat{=}& -\frac{\sqrt{2}}{8} i D \sin\theta \bar\eta_\Delta +
\frac{\sqrt{2} r_\Delta \cot\theta}{4 \Sigma_\Delta} \nn\\
&& -
\frac{\sqrt{2} i a \cos (2\theta)}{4
\sin\theta\Sigma_\Delta},\label{B-theta}\\
 \epsilon&=& \frac{U[(r'-r_\Delta)^2 + a^2 \cos^2\theta + i a \cos\theta r']}{2[(r'-r_\Delta)^2 + a^2 \cos^2\theta]}
  \hat{=} \frac{U}{2},\nn\\
 \c &\hat{=}& -\frac{ i a \cos\theta}{2\Sigma_\Delta}, \c+\bar\c\hat{=}0,\nn\\
\pi&\hat{=}&\a + \bar\b,\nn\\
\Psi_0&\hat{=}&0, \Psi_1 \hat{=} 0, \nn\\
\Im \Psi_2 &\hat{=}& - \frac{i D \cos\theta}{\Sigma_\Delta}(r_\Delta^2 + a^2\cos^2\theta -a^2\sin^2\theta),\nn \\
\Psi_3 &\hat{=} & \frac{i \sqrt{2} \sin\theta r_\Delta \eta_\Delta}{4 \Sigma_\Delta^3}
[D \Sigma_\Delta^2 + 2 i a^2\cos^2\theta],\nn\\
\Psi_4&\hat{=}&0.\nn
 \eea
From this Kerr tetrad in the approximate Bondi
frame, the NP coefficients satisfy
\bea \nu\hat{=}\mu-\bar\mu\hat{=} \pi-\a-\bar\b\hat{=} \c+\bar\c \hat{=}\epsilon-\bar\epsilon\hat{=}0,\nn\\
\pi\hat{=}\bar\tau, \rho\hat{=}\bar\rho, \mu < 0. \eea

\subsection{Bergqvist's calculation\label{bergq}}

After making the null rotation, the complex spatial null tetrad is
\bea m^a = \frac{1}{\sqrt{2}\bar\eta_\Delta} [\partial_\theta
-\frac{i\Sigma_\Delta}{\sin\theta(r_\Delta^2+a^2)} \partial_\phi],
\eea
and the NP $\b$ then becomes \cite{Bergqvist92}
 \bea \b(\theta) = \frac{1}{
2\sqrt{2}\bar\eta_\Delta}[\cot\theta -\frac{2 i a
\sin\theta}{\bar\eta_\Delta} - \frac{i a
(r-M)\sin\theta}{(r^2_\Delta + a^2)}]. \nn\eea
The Dougan-Mason's holomorphic conditions can be written as two
equations in component. The one is
 \bea 0= \eth \lambda_0 =(\d
-\b)\lambda_0,\eea
where the shear vanishes. The solution has the form $\lambda_0 =
C_{-1} A_{-1}(\theta) e^{-i\phi/2} + C_{1} A_{1}(\theta)
e^{i\phi/2} $. The $A_{-1}, A_{1}$ satisfy the differential
equations
\bea \partial_\theta A_{-1} -(G+H) A_{-1} =0,\\
     \partial_\theta A_{1} + (G-H) A_{1} =0.
 \eea
With the aid of using Maple, we can find the solutions are
\bea A_{-1}(\theta)&= e^{\int G+H d \theta} =\frac{
\sin(\theta/2)}{\bar\eta_\Delta} \exp( \frac{i a \cos\theta (r-M-i
a)}{2(r^2_\Delta+a^2)}), \nn\\
 A_{1}(\theta)&= e^{\int -G+H d \theta} = \frac{
\cos(\theta/2)}{\bar\eta_\Delta} \exp( \frac{i a \cos\theta (r-M+i
a)}{2(r^2_\Delta+a^2)}), \nn\eea
where
\bea G&:=& \frac{\Sigma_\Delta}{2
\sin\theta(r^2_\Delta+a^2)},\;\;\;\;
     H := \sqrt{2} \bar\eta_\Delta \b.
\eea
The other holomorphic condition is
\bea 0= \eth\lambda_1 -\mu\lambda_0 =(\d+\b)\lambda_1
-\mu\lambda_0.\eea
 The solution has the form $\lambda_1 = B_{-1}
(\theta) e^{-i\phi/2} + B_{1}(\theta) e^{i\phi/2} $. It satisfies
the differential equations
\bea \partial_\theta B_{-1} -(G-H) B_{-1} -\sqrt{2} \mu \bar\eta_\Delta C_{-1} A_{-1}=0,\\
      \partial_\theta B_{1} + (G+H) B_{1} -\sqrt{2} \mu \bar\eta_\Delta C_{1} A_{1}=0,
 \eea
where
\bea B_1(\theta) = \frac{C_1}{\sqrt{2} A_{-1}} \int h(s) d s, \\
     B_{-1}(\theta) = \frac{C_{-1}}{\sqrt{2} A_{1}} \int h(s) d s,
\eea
and
\bea \int h(\theta) d \theta= \int 2 \mu \eta_\Delta A_1 A_{-1} d
\theta.\eea
Use $\mu= -\frac{r_\Delta}{r_\Delta^2+a^2} -
\frac{(r_\Delta-M)a^2\sin^2\theta}{2(r_\Delta^2+a^2)^2}$, we get
\bea & \int^\pi_0 \mu |A_{-1}|^2 \sin\theta d \theta \nn\\
&=-\frac{\tan^{-1}(\frac{a}{r}) (r^2+a^2)(r+M) - a r (M-r)}{2 a
r(r^2+a^2)^2}, \label{Up1}\eea
and
 \bea &&\int^\pi_0 h(s) d s =\int^\pi_0 2 \mu \eta_\Delta A_1 A_{-1} d
\theta \nn\\
&&= \frac{1}{4}[\sqrt{r^2+a^2} a (2M-2r+1) - 3
r^3\ln(\frac{a+\sqrt{r^2+a^2}}{-a+\sqrt{r^2+a^2}})\nn\\
&&+  2(r+M)a^2\ln(-a+\sqrt{r^2+a^2}) \nn\\
&&- M
r^2\ln\ln(\frac{a+\sqrt{r^2+a^2}}{-a+\sqrt{r^2+a^2}}) - 2 a^2 M \ln a\nn\\
&& -2 r a^2\ln(a+\sqrt{r^2+a^2})]/(a(r^2+a^2)^2).
\label{Down1}\eea
Therefore, we use $-(\ref{Up1})/|\ref{Down1}|$ to calculate
Dougan-Mason mass.Thus we get
 \bea m_{DM} &=& 2 M r_{+} \frac{\int^\pi_0 -\mu(\theta)
|A_{-1} (\theta)|^2 \sin\theta d \theta}{|\int^\pi_0 h(\theta) d
\theta |} \nn\\
&=& M  -\frac{a^2}{24 M}+ O(a^4).\eea
%

\subsection{Using the Dougan-Mason conditions to calculate the Kerr mass in the approximate Bondi
frame} \label{DMKerr-bondi}

The complex spatial null tetrad for Kerr in Bondi coordinate has
the form
 \bea m^a = \frac{1}{\sqrt{2}\eta_\Delta}
(\partial_\theta -\frac{i}{\sin\theta} \partial_\phi). \eea
The NP $\b$ is function of $\theta$ on horizon (recall eq. (\ref{B-theta})) and it has the form
\bea \b(\theta) = -\frac{\sqrt{2}}{8} i D \sin\theta
\bar\eta_\Delta + \frac{\sqrt{2} r_\Delta \cot\theta}{4
\Sigma_\Delta} - \frac{\sqrt{2} i a \cos (2\theta)}{4
\sin\theta\Sigma_\Delta}.\nn\eea
Dougan-Mason's holomorphic conditions in components has two
equations. One is
 \bea 0= \eth \lambda_0 =(\d -\b)\lambda_0.\eea
The solution has the form $\lambda_0 = C_{-1} A_{-1}(\theta)
e^{-i\phi/2} + C_{1} A_{1}(\theta) e^{i\phi/2} $ where $A_1,
A_{-1}$ satisfy
\bea \partial_\theta A_{-1} -(G+H) A_{-1} =0,\;\;
     \partial_\theta A_{1} + (G-H) A_{1} =0.\nn
 \eea
We use Maple to check the solutions are
\bea A_{-1}(\theta)&=& e^{\int G+H d \theta} \nn\\
&=& \sin(\theta/2)
\sqrt{\bar\eta_\Delta} \exp(i\cos\theta D (3 r_\Delta^2 +
a^2\cos^2\theta)/12), \nn\\
 A_{1}(\theta)&=& e^{\int -G+H d \theta} \nn\\
&=&  \cos(\theta/2)
\sqrt{\bar\eta_\Delta} \exp(i\cos\theta D (3 r_\Delta^2 +
a^2\cos^2\theta)/12),\nn \eea
\bea G&:=& \frac{1}{2 \sin\theta}, \;\;\;\;
         H:= \sqrt{2} \eta_\Delta \b.
\eea
The other is
 \bea 0= \eth\lambda_1 -\mu\lambda_0
=(\d+\b)\lambda_1 -\mu\lambda_0.\eea
 The solution has the form $\lambda_1 = B_{-1}
(\theta) e^{-i\phi/2} + B_{1}(\theta) e^{i\phi/2} $.   $B_1,
B_{-1}$ satisfy
\bea \partial_\theta B_{-1} -(G-H) B_{-1} -\sqrt{2} \mu \eta_\Delta C_{-1} A_{-1}=0,\\
      \partial_\theta B_{1} + (G+H) B_{1} -\sqrt{2} \mu \eta_\Delta C_{1}
      A_{1}=0,
 \eea
where
\bea B_1(\theta) = \frac{C_1}{\sqrt{2} A_{-1}} \int h(s) d s, \\
     B_{-1}(\theta) = \frac{C_{-1}}{\sqrt{2} A_{1}} \int h(s) d s,
\eea
and
\bea \int h(\theta) d \theta= \int 2 \mu \eta_\Delta A_1 A_{-1} d
\theta. \eea
For Kerr solution in Bondi coordinate, we have $\mu=
-\frac{r_\Delta}{\Sigma_\Delta}$.
%
%
%
%
%
Thus we get
\bea m_{DM} &=& 2 M r_{+} \frac{\int^\pi_0 -\mu(\theta) |A_{-1}
(\theta)|^2 \sin\theta d \theta}{|\int^\pi_0 h(\theta) d
\theta |} \nn\\
&=& M  -\frac{a^2}{24 M}+ O(a^4)\eea
which is approximate to the Kerr mass when $a$ is very small.

\subsection{Checking our quasi-local formula} \label{myquasi-localKerr}

We replace $\mu_0$ by using $\frac{\Psi_2^0+\eth_0\pi_0 +
\pi_0\bar\pi_0}{2\epsilon_0}$ in order to check our quasi-local
formula. Since we know the NP coefficients and Weyl tensors for
Kerr in Bondi coordinate (See Section \ref{KerrBondi}), we have
\bea \frac{\Psi_2^0+\eth_0\pi_0 + \pi_0\bar\pi_0}{2\epsilon_0}=
-\frac{r_\Delta}{3 \Sigma_\Delta} -\frac{\sin^2\theta
\Sigma_\Delta D^2}{24 U} - \frac{E}{24 U \Sigma_\Delta^3}, \nn\eea
where
\bea E := 8\Sigma_\Delta^2 - 4 r_\Delta^4 + 4 a^2(a^2\cos^2\theta
-r_\Delta^2\sin^2\theta).\eea

%
%
%
%

 We thus get
\bea m_{DM} &=& 2 M r_{+} \frac{\int^\pi_0
-[\frac{\Psi_2^0+\eth_0\pi_0 + \pi_0\bar\pi_0}{2\epsilon_0}]
|A_{-1} (\theta)|^2 \sin\theta d \theta}{|\int^\pi_0 h(\theta) d
\theta |} \nn\\
&=& M  -\frac{a^2}{24 M}+ O(a^4),\eea
which is approximate to the Kerr mass when angular momentum per
mass is very small.

\textbf{Remark.}
We conclude that the quasi-local mass for these three cases are
the same up to $a^2$. This is not surprising because they are based on
 Dougan-Mason's quasi-local mass definition. Even
though our work for the Kerr in Bondi coordinate is just
approximation, we still can get the same quasi-local mass up to
the second order. However, if one compare with the Hawking mass,
the second order is different from the Dougan-Mason mass.
 Hawking mass of Kerr horizon is
\bea m_H &=& M [\frac{1+
(1-a^2/M^2)^\half}{2}]^\half \nn\\
&=& M - \frac{a^2}{8 M} + O(a^4). \eea
%

\section{Quasi-local mass and energy flux for the Vaidya
solution} \label{qlVaidya}

A good example to study the dynamical horizon is Vaidya solution.
For this, we know that the dynamical horizon is a space-like
hypersurface if the black hole is growing. On the contrary, if
black hole is contracting then it's a time-like hypersurface. In
the later case the dominate energy condition will not hold  so we
restrict attention to the former case.  The outgoing null normal
of the Vaidya solution is $\l^a =\frac{\partial}{\partial v} +(
\frac{r'}{2(r'+r_\Delta)}-\dot r_\Delta) \frac{\partial}{\partial
r'}$, we also have $\rho\hat{=}\sigma\hat{=} 0$ on horizon, and
$\Phi_{00} =\frac{\dot M}{r^2}$. We now show that if the dominate
energy condition is satisfied then $\dot r_\Delta >0$. From the
Raychaudhuri equation $\L_\l \Theta_{(\l)} = - \Theta_{(\l)}^2
-\sigma_{ab}\sigma^{ab} +
\omega_{ab}\omega^{ab}-2\epsilon\Theta_{(\l)} +\half R_{ab}\l^a
\l^b$. The expansion of the outgoing null normal of Vaidya is
$\Theta_{(\l)} = - Re{\rho} = \frac{r-2 M(v)}{2
r^2}\hat{=}\frac{r'}{2 r_\Delta^2}$, so it vanishes on horizon.
However, it's time derivative does not vanish. Here we have $\L_\l
\Theta_{(\l)}\hat{=} -\frac{\dot r_\Delta}{2 r_\Delta^2}\hat{=} -
\Phi_{00} \leq 0$. If the dominate energy condition satisfied,
then $\Phi_{00}\geq 0$, i.e., $\dot r_\Delta \geq 0$. Let $V$
denote a vector field which is tangential to the dynamical horizon
$H^+$, is everywhere orthogonal to the foliation by marginally
trapped surfaces and preserves this foliation. We can always
choose the normalization of $\l^a$ and $n^a$ such that $\l^a n_a
=1$ and $V^a = \l^a - f n^a$ for some $f$. Since $V^a V_a = -2 f
$, it follows that $H^+$ is space-like, null or time-like,
depending on whether $f$ is positive, zero or negative. We now
show that if the dominate energy condition holds then  $f$ is
non-negative. Let us begin by noting that the definition of $V^a$
immediately implies $\L_V \Theta_{(\l)} =0$, whence, $\L_\l
\Theta_{(\l)}= f \L_n \Theta_{(\l)}$. Therefore, the Raychaudhuri
equation for $\l^a$ implies
$ f \L_n \Theta_{(\l)} = - |\sigma|^2 -\Phi_{00} \leq 0. $
Physically, we can assume that the expansion of the outgoing null
normal becomes negative if it moves along the incoming direction
to the interior of the marginally trapping surfaces, so $\L_n
\Theta_{(\l)} <0$. Hence, if the dominate energy condition is
satisfied, then $f$ is non-negative, i.e., the dynamical horizon
$H^+$ is space-like or null. The Vaidya solution is similar with
the Schwarzschild solution but the mass is time dependent and it's
a non-vacuum dust solution. In this appendix we calculate our
quasi-local mass and flux for the Vaidya solution and compare with
the results one expects from the form of Vaidya metric. In
advanced Eddington-Finkelstein coordinates $(v, r , \theta,
\phi)$,
%
%
the metric is
\be\ba{l} d s^2 = (1-\frac{2 M(v)}{r}) d v^2- 2 d v d r -r^2 (d
\theta^2 + \sin^2\theta d \phi^2).\ea\ee
where the horizon radius $r_\Delta(v)= 2 M(v)$ is time dependent.
When $r$ approaches the horizon radius $r_\Delta(v)$, the metric
approaches to
\be\ba{l} d s^2 \hat{=} -2 \dot r_\Delta d v^2 - r_\Delta^2(d
\theta^2 + \sin^2\theta d \phi^2) \ea\ee
where we use a coordinate transformation $d r' = d r - \dot
r_\Delta d v$. It's obvious that the Vaidya horizon is space-like
or null hypersurface since it has the metric signature $(---)$ if
$\dot r_\Delta\geq 0$. It's null when the mass approach to
constant and it returns to the Schwarzschild solution.
In the new coordinate $(v, r' , \theta, \phi)$, the tetrad
components  are
\be\ba{lllllllll} \l^a &=& (1 &,&  \frac{r'}{2 (r'+
r_\Delta)}-\dot r_\Delta &,&
0 &,&  0)\\
n^a &=& (0 &,& -1 &,& 0 &,& 0)\\
m^a &=& (0 &,& 0  &,& \frac{1}{\sqrt{2} (r'+r_\Delta)} &, &
-\frac{i}{\sqrt{2}  (r'+r_\Delta)\sin\theta})
 \ea\ee
The NP coefficients are
\be\ba{l} \kappa=\sigma=\lambda=\nu=\tau=\pi=\gamma=0,\\
\rho= -\frac{r-2 M(v)}{2 r^2}=\frac{r'}{2 r_\Delta^2}-
\frac{r'^2}{r_\Delta^3}+ O(r'^3),\\
 \mu=-\frac{1}{r}= -\frac{1}{r_\Delta}+\frac{r'}{r_\Delta^2}+O(r'^2) ,\\
 \epsilon=\frac{M(v)}{2
r^2}=\frac{1}{4 r_\Delta}+O(r'),\\
\a= -\b= -\frac{\sqrt{2} \cot\theta}{4 r_\Delta}+O(r'),\\
\Psi_0=\Psi_1=\Psi_3=\Psi_4=0, \Psi_2=-\frac{M(v)}{r_\Delta^3}+O(r'), \\
\Phi_{00}= \frac{\dot M(v) }{r_\Delta^2}+O(r'). \ea\ee

\textbf{Quasi-local mass of Vaidya horizon}

Let's calculate the horizon energy momentum near the Vaidya
horizon and examine the related mass gain or loss formula near
such dynamical horizon. We use the holomorphic gauge fixing, so
the horizon quasi-local energy-momentum is
\bea P^{\underline{A}\underline{A'}} = -\frac{1}{4\pi} \oint_S \mu
\lambda^{\underline{A}}_0 \lambda^{\underline{A'}}_{0'} d S.\eea
Here we use the Dougan-Mason definition of mass:
\bea m^2_{DM} = P_{\underline{AA'}} P^{\underline{AA'}}=
\frac{2}{|N|^2}
(P^{\underline{00'}}P^{\underline{11'}}-P^{\underline{01'}}P^{\underline{10'}})\eea
where $N=\varepsilon^{\underline{01}}=
\lambda^{\underline{0}}_0\lambda^{\underline{1}}_1-\lambda^{\underline{0}}_1\lambda^{\underline{1}}_0$.
To calculate the Dougan-Mason mass, we have to solve the
holomorphic equations for the constant spinors equations. Here we
simply follow the calculation that done by Bergvist
\cite{Bergqvist92}. The holomorphic conditions:
\bea \eth \lambda_0 +\sigma \lambda_1 =0 \Rightarrow &
(\delta-\b)\lambda_0 =0 \\
\eth \lambda_1 -\mu\lambda_0 =0  \Rightarrow &  (\delta +\b)
\lambda_1 -\mu\lambda_0 =0 \eea
For $\eth \lambda_0=0$, we have to solve the PDE
\bea (\frac{\partial}{\partial \theta}
-\frac{i}{\sin\theta}\frac{\partial}{\partial \phi} -
\frac{\cot\theta}{2})\lambda_0 =0, \eea
then we get the solution
$ \lambda_0 =C_{-1} \lambda^{\underline{0}}_0 + C_1
\lambda^{\underline{1}}_0. $
For $\eth \lambda_1 -\mu\lambda_0 =0$, we have to solve the PDE
\bea
 \frac{1}{\sqrt{2}r_\Delta}
[(\frac{\partial}{\partial \theta}
-\frac{i}{\sin\theta}\frac{\partial}{\partial \phi} +
\frac{\cot\theta}{2})\lambda_1+\sqrt{2} \lambda_0]=0, \eea
then we get
$ \lambda_1 =C_{-1} \lambda^{\underline{0}}_1 + C_1
\lambda^{\underline{1}}_1 .$
Here
\be\ba{lllll} \lambda^{\underline{0}}_0 &=& A_{-1}e^{-i
\phi/2},\;\;
     \lambda^{\underline{1}}_0 = A_1 e^{i \phi/2},\\
     \lambda^{\underline{0}}_1 &=& \frac{\int_\pi^\theta B(s) d s \; e^{-i \phi/2}}{\sqrt{2}
     A_1(\theta)},\;\;
     \lambda^{\underline{1}}_1 = \frac{\int_0^\theta B(s) d s \; e^{i \phi/2}}{\sqrt{2}
     A_{-1}(\theta)},\\
     A_1 &=& \cos(\theta/2)/r,\;\;
     A_{-1} =  \sin(\theta/2)/r,\\
     B(\theta) &=& \sin\theta /r^2.
\ea\ee
We choose two independent solution $C_{-1} =1, C_1 =0$ for
$\lambda^{\underline{0}}_A$ and $C_{-1} =0, C_1 =1$ for
$\lambda^{\underline{1}}_A$ here. So when we integrate over
$\phi$, $\phi\leq \phi \leq 2\pi$ in $P^{\underline{AA'}}$, we
find $P^{\underline{01'}} = P^{\underline{10'}}$ and
$P^{\underline{00'}}=P^{\underline{11'}}$. Therefore, the
Dougan-Mason mass becomes
\bea m_{DM} = \frac{\sqrt{2} P^{\underline{00'}}}{|N|} \eea
where
\bea P^{\underline{00'}} &=& -\frac{1}{4\pi} \int \mu |A_{-1}|^2 d S^2
\hat{=} \frac{1}{2 r_\Delta} \eea
and
$ N
=\lambda^{\underline{0}}_0\lambda^{\underline{1}}_1-\lambda^{\underline{0}}_1\lambda^{\underline{1}}_0
 = \frac{1}{\sqrt{2}} \int_0^\pi h(\theta) d \theta
 = -\frac{\sqrt{2}}{r^2}.$
Finally, the Vaidya horizon mass from the Dougan-Mason's mass
definition is
\bea m_{DM} = \frac{r_\Delta}{2}=M(v)\eea
which agrees with the expected result.

\textbf{Quasi-local flux of Vaidya horizon}

Next, we reduce our flux formula to Vaidya by taking shear
vanishing and compare it by using the solution of the holomorphic
constant spinors. For Vaidya solution, we have
\bea\mu_0 = -\frac{1}{r_\Delta(v)}, \sigma'_0=  \sigma_0=0\eea
where $r_\Delta$ is time dependent. We substitute these formulae
for the NP quantities from the Vaidya solution into our
quasi-local energy-momentum integral we obtain in
(\ref{ch5-quasi-local}). Then
\bea I(r_\Delta) &=& - \frac{1}{4\pi} \oint \mu_0 \lambda_0^0
\bar\lambda_{0'}^0 d
S_\Delta \label{I1}\\
&=& - \frac{1}{4\pi}\oint \frac{\Psi^0_2}{2 \epsilon_0}
\lambda_0^0 \bar\lambda_{0'}^0 d S_\Delta. \label{I2} \eea
Then we rescale $\lambda_0^0$ so that we get the condition $\dot
\lambda^0_0 =0$ (recall (\ref{TCDdhs})) from our asymptotic
constant spinor condition for the dynamical horizon. We use the
fact that $\frac{\partial }{\partial v}d S_\Delta = 2\frac{\dot
r_\Delta}{r_\Delta} d S_\Delta$.

From the first expression (\ref{I1}), the quasi-local energy flux
is
\bea \frac{d I}{d v} &=& - \frac{1}{4\pi} \oint(\dot\mu_0 + 2
\mu_0 \frac{\dot
r_\Delta}{r_\Delta}) \lambda_0^0 \bar\lambda_{0'}^0 d S_\Delta \\
&=& \frac{1}{4\pi} \oint \frac{\dot M}{2 M^2}\lambda_0^0
\bar\lambda_{0'}^0 d S_\Delta. \eea
From the second expression (\ref{I2}), it becomes
 \bea \frac{d I}{d v} &=&
- \frac{1}{4\pi} \oint(\frac{\dot\Psi^0_2}{2\epsilon_0} -
\frac{\Psi^0_2 \dot\epsilon_0}{2 \epsilon_0^2} + \frac{\Psi^0_2
\dot r_\Delta}{\epsilon_0 r_\Delta}) \lambda_0^0 \bar\lambda_{0'}^0 d S_\Delta \nn\\
&&\textrm{(Use  $\Psi^0_2 = - \frac{1}{8 M^2}, \epsilon_0
=\frac{1}{8M} $)}\nn\\
&=&  \frac{1}{4\pi} \oint \frac{\dot M}{2 M^2}\lambda_0^0
\bar\lambda_{0'}^0 d S_\Delta. \nn\eea
We use the Dougan-Mason's mass definition, then  $\dot m_{DM} =
\frac{\sqrt{2} \dot P^{\underline{00'}}}{|N|}$. Therefore we can
put $\lambda^0_0 = \lambda^{\underline{0}}_0 $ in our quasi-local
formula. The Vaidya's quasi-local mass flux is
 \bea \frac{d
m_{DM}}{d v}& =& \frac{r^2}{4\pi} \int\frac{ \dot M}{2
M^2} \frac{\sin(\theta/2)^2}{r^2} d S
=\dot M.\eea
If the dominate energy condition satisfied, i.e., $\dot M \geq
0$., then we find the mass gain law for the Vaidya. 

\end{appendix}

\end{document}